\documentclass{nature}
\usepackage[super,numbers,sort&compress]{natbib}
\usepackage{graphicx}  
\usepackage{dcolumn}   
\usepackage{bm}        
\usepackage{verbatim}   
\usepackage{graphicx}
\usepackage{epstopdf}
\usepackage{footnote}
\usepackage{epsfig}
\usepackage{subfigure}
\usepackage{amssymb}
\usepackage{amsmath,bm}
\usepackage{xcolor}
\usepackage{float}
\usepackage{subfigure}
\usepackage[most]{tcolorbox}
\usepackage{soul}
\usepackage{color}
\usepackage[framemethod=TikZ]{mdframed}
\usepackage{lipsum}
\mdfdefinestyle{MyFrame}{%
	linecolor=blue,
	outerlinewidth=2pt,
	roundcorner=20pt,
	innertopmargin=\baselineskip,
	innerbottommargin=\baselineskip,
	innerrightmargin=20pt,
	innerleftmargin=20pt,
	backgroundcolor=gray!50!white}

\title{Full-Duplex Nonreciprocal-Beam-Steering Metasurfaces Comprising Time-Modulated Twin Meta-Atoms}
\author{Sajjad Taravati $^{1}$ and George V. Eleftheriades $^{1}$}	
\makeatletter
\let\saved@includegraphics\includegraphics
\AtBeginDocument{\let\includegraphics\saved@includegraphics}
\renewenvironment*{figure}{\@float{figure}}{\end@float}
\makeatother
\begin{document}
	\maketitle
\begin{affiliations}
	\item The Edward S. Rogers Sr. Department of Electrical and Computer Engineering, University of Toronto, Toronto, Ontario M5S 3G4, Canada\\
	Email: sajjad.taravati@utoronto.ca
\end{affiliations}
	
\begin{abstract}
We present the concept, theoretical model and experimental implementation of a full-duplex nonreciprocal-beam-steering transmissive phase-gradient metasurface. Such a metasurface is realized by exploiting the unique properties of the frequency-phase transition in coupled time-modulated twin meta-atoms. The metasurface may be placed on top of a source antenna to transform the radiation pattern of the source antenna, and introduce different radiation patterns for the transmit and receive states. In contrast to the recently proposed applications of time modulation, here the incident and transmitted waves share the same frequency. The metasurface is endowed with directive, diverse and asymmetric transmission and reception radiation beams, and tunable beam shapes. Furthermore, these beams can be steered by simply changing the modulation phase. The proposed coupled meta-atoms inherently suppress undesired time harmonics, leading to a high conversion efficiency which is of paramount importance for practical applications such as point to point full-duplex communications.
\end{abstract}

\section{Introduction}
The ever increasing progress in wireless telecommunication systems demands a new class of compact structures, called metasurfaces, for wave engineering and controlling the electromagnetic wave radiation ~\cite{Alu_PRB_2015,Taravati_2016_NR_Nongyro,Fan_APL_2016,wang2018extreme,zhang2019breaking,Grbic2019serrodyne}. Even though conventional reciprocal and static metasurfaces are capable of providing quite useful operations, nonreciprocity and time modulation can bring metasurfaces to a whole new level, introducing peculiar and unique wave engineering functionalities not seen in conventional metamaterials and metasurfaces~\cite{Taravati_2016_NR_Nongyro,sounas2017non,zang2019nonreciprocal,Taravati_Kishk_TAP_2019,ptitcyn2019time}. 

Nonreciprocity based on time modulation represents a powerful tool for advanced wave engineering and extraordinary control over electromagnetic waves~\cite{cassedy1965waves,Cassedy_PIEEE_1967,sounas2017non,Taravati_PRAp_2018,Taravati_Kishk_TAP_2019,inampudi2019rigorous,elnaggar2019generalized,Taravati_Kishk_PRB_2018,wang2018photonic,Taravati_Kishk_MicMag_2019}. Recently, there has been a deep investigation on wave propagation and scattering in time-periodic~\cite{zurita2009reflection,martinez2018parametric,salary2018time,mirmoosa2019time,ptitcyn2019time}, and space-time-periodic~\cite{Fan_NPH_2009,Taravati_PRAp_2018,elnaggar2019generalized,li2019nonreciprocal,oudich2019space,correas2018magnetic,liu2018huygens,du2019simulation} media. Time modulation represents a thrilling topic thanks to the complexity and rich physics of the problem, as well as the diverse and exciting practical applications of time-modulation. Time modulation has been recently used for the realization of isolators and nonreciprocal platforms~\cite{wentz1966nonreciprocal,Fan_NPH_2009,Fan_PRL_109_2012,Wang_TMTT_2014,zanjani2014one,Bahl_2015non,Taravati_PRB_2017,Taravati_PRB_SB_2017,Bahl_2018time,correas2019plasmonic}, circulators~\cite{wentz1966nonreciprocal,estep2014magnetic,reiskarimian2016magnetic}, metasurfaces~\cite{Alu_PRB_2015,Fan_APL_2016,Fan_mats_2017,Salary_2018,salary2019dynamically,Taravati_Kishk_TAP_2019,zhang2019breaking,zang2019nonreciprocal_metas}, frequency converters~\cite{Taravati_PRB_Mixer_2018,Grbic2019serrodyne}, mixer-duplexer-antennas~\cite{Taravati_LWA_2017}, antennas~\cite{shanks1961new,Alu_PNAS_2016,ramaccia2018nonreciprocity,salary2019nonreciprocal,zang2019nonreciprocal,taravati2018space,taravati2019_mix_ant}, unidirectional beam splitters~\cite{Taravati_Kishk_PRB_2018}, nonreciprocal filters ~\cite{alvarez2019coupling,wu2019isolating} and impedance matching structures~\cite{shlivinski2018beyond}. Other recently reported outstanding and unique properties and applications of space-time-periodic media include anomalous topological edge states in space-time photonic crystal~\cite{oudich2019space}, Fresnel drag~\cite{huidobro2019fresnel}, signal coding metagratings and metasurfaces~\cite{zhang2018space,taravati_PRApp_2019}.
	
Previously reported time-modulated metasurfaces suffer from an undesired frequency alteration in the spectrum of the incident wave. Such a frequency change is usually undesired as the frequency of the up-converted/down-converted wave is very close to the incident wave and cannot be represented as an effective mixing functionality. In addition, the converted wave is usually accompanied by an infinite number of higher order time harmonics, which leads to a crowded frequency spectrum, poor conversion efficiency, and may result in inter-modulation interference. Another important point is that most of the previously reported time-modulated metasurfaces are reflective structures~\cite{Alu_PRB_2015,Fan_APL_2016,Fan_mats_2017,Salary_2018,salary2019dynamically,zang2019nonreciprocal}, which may not be as practical as transmissive metasurfaces. 
	
This study presents the concept, theoretical model and experimental implications of full-duplex nonreciprocal-beam-steering transmissive phase-gradient metasurfaces. Such metasurfaces may be placed on top of a source antenna, transforming their radiation pattern of the source antenna and providing different radiation patterns for the transmit and receive states. Such metasurfaces are composed of an array of coupled time-modulated twin meta-atoms, each of which functioning four major operations, i.e. wave reception, nonreciprocal phase shift for nonreciprocal beam steering, filtering out of unwanted temporal harmonics, and wave radiation. 

The contribution of this work is as follows. We propose a metasurface which can be used for multi-functionality purposes, i.e. different radiation patterns (e.g., different half power beamwidths and different angle of radiation, as well as, different signal amplifications) for transmission and reception, nonreciprocal-beam-steering and beam shaping. Such a metasurface provides highly directive and tunable beams for point to point full-duplex telecommunications. In addition, from the physical point of view, this study presents the first time-modulated antenna metasurface platform, where the incident and transmitted waves share the same temporal frequency, different than other previously reported time-modulated radiating metasurfaces. We propose the concept of coupled time-modulated twin meta-atoms, and take advantage of the physical phenomena that occur in such structures. Specifically, the nonreciprocal phase shift in a round-trip photonic transitions in periodic time-modulated media with zero frequency alteration, transmissive reception and re-radiation of the electromagnetic waves, as well as, filtering of undesired temporal harmonics. The proposed coupled time-modulated meta-atoms inherently prohibits the excitation of undesired time harmonics, leading to a high conversion efficiency which is of paramount importance for practical applications such as for instance point to point full-duplex telecommunications.

Firstly, we present the concept of full-duplex nonreciprocal-beam-steering transmissive metasurface. Subsequently, we present the theoretical, physical and practical implications of nonreciprocal phase shift in coupled time-modulated twin meta-atoms. We next propose the implementation scenario for the practical realization of a full-Duplex nonreciprocal-beam-steering transmissive metasurface induced by phase-gradient time-modulated twin meta-atoms. Subsequently, we present full-wave simulation, as well as, experimental results for symmetric and asymmetric radiation beams at different angles of radiation.

\section{Concept of Nonreciprocal Radiation Beam}\label{sec:concept}
\begin{figure*}[h!]
	\begin{center}
		\includegraphics[width=0.8\columnwidth]{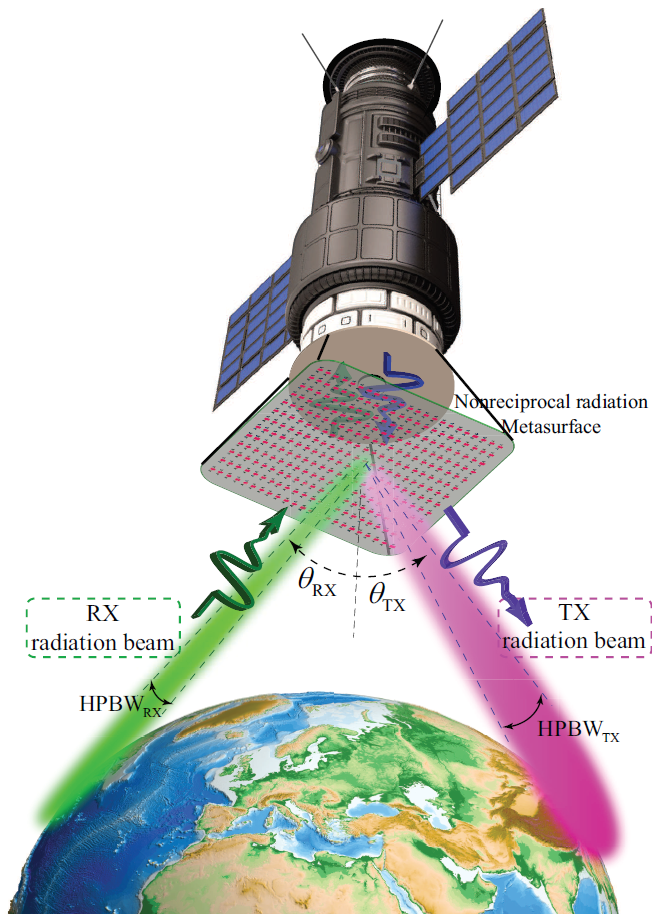}
		\caption{Full-duplex nonreciprocal radiation beam yielding highly directive reception and transmission radiation beams for efficient full-duplex point to point telecommunications.}
		\label{Fig:sat}
	\end{center}
\end{figure*}

Figure~\ref{Fig:sat} illustrates the functionality of the nonreciprocal radiation beam from a gradient metasurface for efficient full-duplex point to point telecommunications. In the transmission state, the wave is launched from the source antenna traveling along the $+z$ direction, passes through the metasurface from region~1 to region~2 and radiates at angle $\theta_\text{2,TX}$. In contrast, in the receive state, the metasurface presents the maximum transmission from region~2 to region~1 for the incoming wave at angle $\theta_\text{2,RX}$. Therefore, for a given radiation angle $\theta_0$, the metasurface is nonreciprocal, and may be represented by asymmetric and nonreciprocal radiation beams, as
\begin{equation}
F_\text{TX} (\theta)\ne F_\text{RX}(\theta),
\end{equation}
where $F_\text{TX} (\theta)= E_{\theta,\text{TX}} /E_{\theta,\text{TX}}(\text{max})$ and $F_\text{RX}(\theta)=E_{\theta,\text{RX}}/E_{\theta,\text{RX}}(\text{max})$.

\begin{figure}[h!]
	\begin{center}
		\includegraphics[width=0.8\columnwidth]{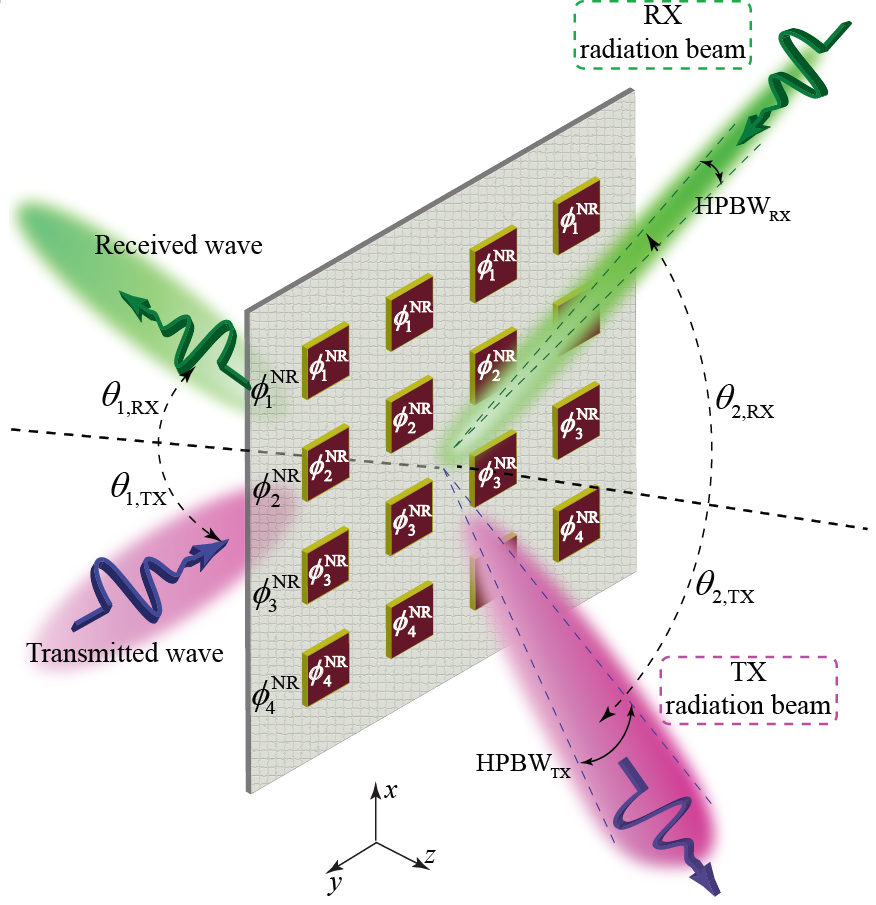}
		\caption{Schematic representation of the gradient metasurface formed by nonreciprocal phase shift meta-atoms.}
		\label{Fig:metas_st}
	\end{center}
\end{figure}

To realize the full-duplex nonreciprocal-beam-steering radome in Fig.~\ref{Fig:sat}, we consider a transmissive metasurface formed by an array of meta-atoms. Figure~\ref{Fig:metas_st} depicts the structure of the metasurface sandwiched between two semi-infinite regions, i.e., region 1 and region 2. Qualitative radiation beams for the transmission and reception states at the two sides of the metasurface are sketched to show the operation principle of the structure. In the transmission state, a plane wave with frequency $\omega_\text{i}$ impinges on the metasurface from the bottom left side with an angle of incidence $\theta_\text{1,TX}$. The
outgoing wave at $\theta_\text{2,TX}$ acquires a discrete phase-profile $\phi(x)$, 
\begin{equation}\label{eq:phi_tx}
\phi (md)=\phi_m,
\end{equation}
where $m$ is the modulation phase at the $m$th meta-atom and $d$ is the spacing between each two adjacent meta-atoms. However, in the transmission state, considering the scheme in Fig.~\ref{Fig:metas_st}, a plane wave with frequency $\omega_\text{i}$ impinges on the metasurface from the top right side with an angle of incidence $\theta_\text{2,RX}$. In contrast to the transmission state, and due to the conservation of momentum, the
outgoing wave at $\theta_\text{1,RX}$ acquires a discrete phase-profile $\phi(x)$, where
\begin{equation}\label{eq:phi_rx}
-\phi (md)=-\phi_m.
\end{equation}

Assuming a constant gradient phase shift along the metasurface, the generalized Snell’s law of refraction yields
\begin{equation}\label{eq:tx}
\frac{\partial \phi_\text{TX}}{\partial x}=k_2 \sin(\theta_\text{2,TX}) -k_1\sin(\theta_\text{1,TX}),
\end{equation}
for the transmission state, and
\begin{equation}\label{eq:rx}
\frac{\partial \phi_\text{RX}}{\partial x}=k_2\sin(\theta_\text{2,RX})-k_1 \sin(\theta_\text{1,RX}) ,
\end{equation}
for the reception state. Here, $k_\text{1}$ and $k_\text{2}$ are the wave numbers in region 1 and region 2, respectively. Considering a constant phase gradient $\partial \phi_\text{MS}/\partial x$, the outgoing wave acquires anomalous refraction with respect to the incident
wave, whereas a spatially variant gradient, i.e, $\partial \phi_\text{MS}/\partial x$, leads to arbitrary radiation beams which enables beam-forming and advanced beam steering purposes.

\section{Theoretical Implications}\label{sec:impl}
\subsection{Theory of nonreciprocal phase shift based on time modulation}
\begin{figure}
	\begin{center}
			\includegraphics[width=0.99\columnwidth]{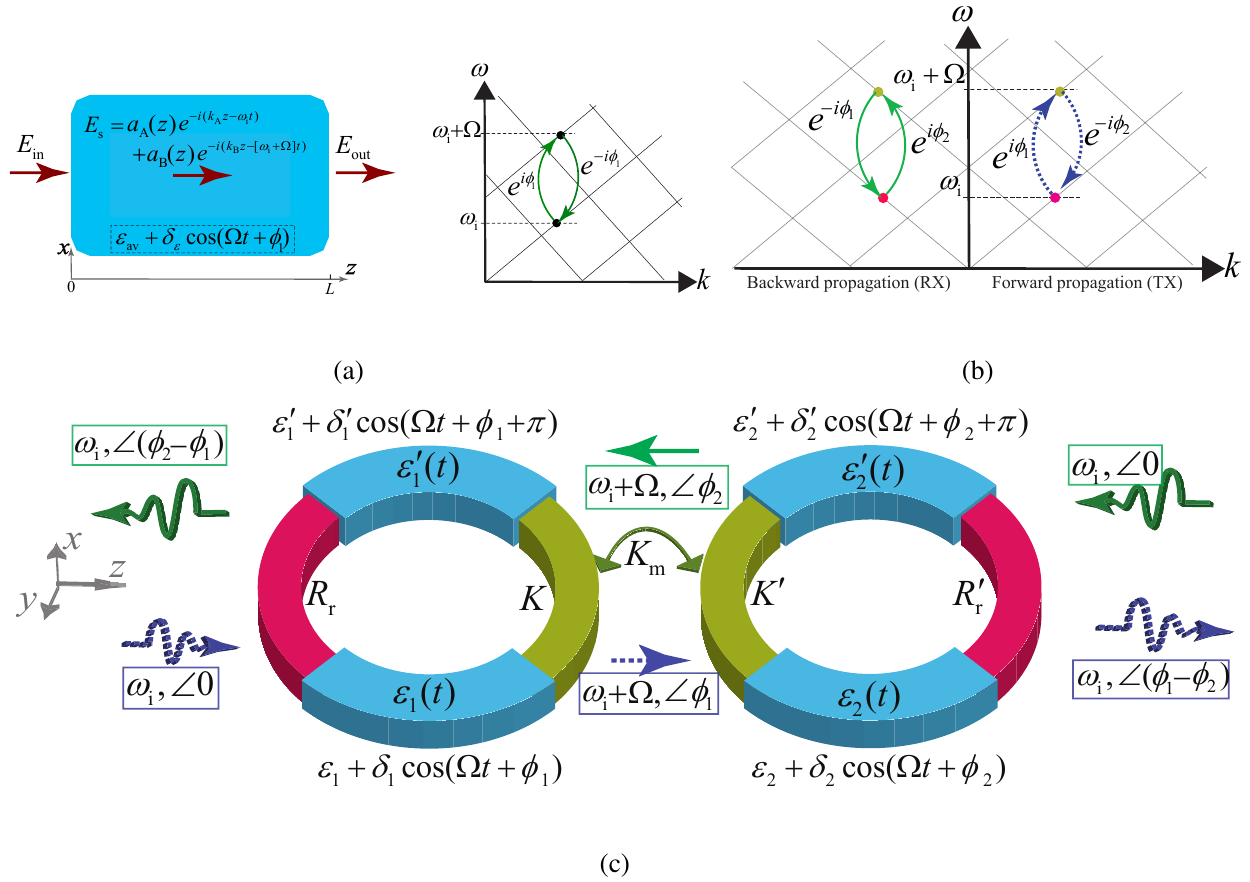}
		\caption{Nonreciprocal phase shifted radiation based on photonic transitions in time-modulated media. (a) (left) A time-modulated medium with permittivity in Eq.~\eqref{eqa:perm}. (right) Dispersion diagram demonstrating the general concept of nonreciprocal phase shift based on photonic transitions in a time-modulated meta-atom supporting two resonant frequencies, $\omega_\text{i}$ and $\omega_\text{i}+\Omega$, and phase shift $\phi_1$. Each arrow represents a photonic transition resulting in a temporal frequency shift accompanied with a phase shift. (b) Dispersion diagram demonstrating nonreciprocal phase shift based on photonic transitions in two time-modulated coupled meta-atoms supporting two resonant frequencies, $\omega_\text{i}$ and $\omega_\text{i}+\Omega$, and two phase shifts $\phi_1$ and $\phi_2$. (c) Coupled twin time-modulated radiating meta-atoms introducing nonreciprocal phase shift for the radiated waves.}
		\label{Fig:coup}
	\end{center}
\end{figure}

Figure~\ref{Fig:coup}(a) (left) sketches the wave propagation and transmission in a periodic time-modulated medium, characterized with a time-dependent permittivity, as 
\begin{equation}\label{eqa:perm}
\epsilon(t)=\epsilon_\text{av} + \delta_{\epsilon} \cos(\Omega t+\phi_1).
\end{equation}
where $\epsilon_\text{av}$ is the average permittivity of the background medium, $\delta_{\epsilon}$ is the modulation amplitude, $\Omega$ denotes the modulation frequency, and $\phi_1$ represents the modulation phase. The right plot in Fig.~\ref{Fig:coup}(a) (right) shows a qualitative dispersion diagram of such a time-modulated periodic medium and photonic transitions between two supported resonant frequencies, $\omega_\text{i}$ and $\omega_\text{i}+\Omega$.

The electric field inside the medium is defined based on the superposition of two supported space-time harmonic fields, i.e.,
\begin{equation}\label{eqa:el}
E_\text{s}(x,z,t)=a_\text{A}(z) e^{-i \left(k_\text{A} z -\omega_\text{i} t \right)}+a_\text{B}(z) e^{-i \left(k_\text{B} z -(\omega_\text{i}+\Omega) t \right)},
\end{equation}
We consider $\Omega<< \omega_\text{i}$. The corresponding wave equation reads $c^2 \partial^{2} \textbf{E}/ \partial z^{2}= \partial^{2} [\epsilon_\text{eq}(t) \textbf{E}]/\partial t^{2}$. Inserting the electric field in~\eqref{eqa:el} into the wave equation results in 
\begin{equation}
\begin{split}
&\left( \frac{\partial^{2} }{\partial z^{2}}  \right) \left[  a_\text{A}(z) e^{-i \left(k_\text{A} z -\omega_\text{i} t \right)}+a_\text{B}(z) e^{-i \left( k_\text{A}z -(\omega_\text{i}+\Omega) t \right)}   \right]\\
&   = \frac{1}{c^2} \frac{\partial^{2} }{\partial t^{2}} 
\left(\bigg[\epsilon_\text{av} + \frac{\delta}{2} e^{i(\Omega t+\phi_1)}  + \frac{\delta}{2} e^{-i(\Omega t+\phi_1)} \right] \left(a_\text{A}(z) e^{-i \left(k_\text{A} z -\omega_\text{i} t \right)}+a_\text{B}(z) e^{-i \left( k_\text{A}z -(\omega_\text{i}+\Omega) t \right)} \right) \bigg),
\end{split}
\end{equation}
and applying the space and time derivatives, while using a slowly varying
envelope approximation, multiply both sides with $e^{i \left(k_\text{A} z -\omega_\text{i} t \right)}$, which gives
\begin{equation}\label{eqa:eq2}
\begin{split}
& \left[k_\text{A}^2 a_\text{A}(z) -2ik_\text{A} \frac{\partial a_\text{A}(z)}{\partial z} \right] +\left[ k_\text{A}^2  a_\text{B}(z) -2i(k_\text{A}) \frac{\partial a_\text{B}(z)}{\partial z} \right] e^{i \Omega t }\\
& \qquad = \frac{1}{c^2} 
\left(\bigg[\omega_\text{i}^2\epsilon_\text{av}+ (\omega_\text{i}+\Omega)^2 \frac{\delta}{2} e^{i(\Omega t+\phi_1)}  + (\omega_\text{i}-\Omega)^2\frac{\delta}{2} e^{-i(\Omega t+\phi_1)} \right] a_\text{A}(z) \\
&\qquad+\left[\omega_\text{i}^2 \epsilon_\text{av} + (\omega_\text{i}+2\Omega)^2\frac{\delta}{2}  e^{i(\Omega t+\phi_1)}  + \omega_\text{i}^2 \frac{\delta}{2}  e^{-i(\Omega t+\phi_1)} \right] a_\text{B}(z) e^{i \Omega t }  \bigg),
\end{split}
\end{equation}
and next, applying $\int_{0}^{\frac{2\pi}{\Omega}}dt$ to both sides gives
\begin{subequations}
\begin{equation}\label{eqa:eq222}
\begin{split}
\frac{d a_\text{A}(z)}{d z}   =  \frac{ik_0^2 \delta}{4 k_\text{A}}  e^{-i\phi_1} a_\text{B}(z)   ,
\end{split}
\end{equation}

Following the same procedure, we next multiply both sides with $e^{-i \Omega t }$, and applying $\int_{0}^{\frac{2\pi}{\Omega}}dt$ in both sides of the resultant, and applying $\int_{0}^{\frac{2\pi}{\Omega}}dt$ to both sides of~\eqref{eqa:eq2} yields
\begin{equation}\label{eqa:eq222b}
\begin{split}
\frac{d a_\text{B}(z)}{d z}   =  \frac{ik_{0}^2 \delta}{4 k_\text{A} }  e^{i\phi_1} a_\text{A}(z)   ,
\end{split}
\end{equation}
\end{subequations}
where $k_{0}=(\omega_\text{i}+\Omega)/c^2$. Equations~\eqref{eqa:eq222} and ~\eqref{eqa:eq222b} represent the coupled-wave equation of the periodic time-modulated medium in Fig.~\ref{Fig:coup}(a).

\subsubsection{Application of boundary conditions:} 
(i) Considering incidence of a wave with frequency $\omega_\text{i}$ (up-conversion corresponding to the left arrow in the dispersion diagram in the right side of Fig.~\ref{Fig:coup}(a)), i.e., $E_\text{in}=E_\text{s}(z=0)=E_0 e^{i \omega_\text{i} t }$, gives $a_\text{A}(z)=E_0 \cos\left( \delta k_1 z /4 \right)$, and $a_\text{B}(z)=i E_0 \sin\left( \delta k_1 z /4 \right)  e^{i \phi_1}$. Considering the coherence length of $L_\text{c}=2\pi k_1/\delta$, $a_\text{A}(z=L_\text{c})=0$ and $a_\text{B}(z=L_\text{c})= i E_0 e^{i \phi_1}$
\begin{equation}\label{eqa:up}
E_\text{out,up-c}=i E_0 e^{-i \left(k_\text{B} z -(\omega_\text{i}+\Omega) t \right)} e^{i \phi_1},
\end{equation}

(ii) Considering incidence of a wave with frequency $\omega_\text{i}+\Omega$ (down-conversion corresponding to the right arrow in the dispersion diagram in the right side of Fig.~\ref{Fig:coup}(a)), i.e., $E_\text{in}=E_\text{s}(z=0)=E_0 e^{i (\omega_\text{i}+\Omega) t }$, gives $a_\text{A}(z)=i E_0 \sin\left( \delta k_0 z /4 \right)  e^{-i \phi_1}$, and $a_\text{B}(z)= E_0 \cos\left( \delta k_0 z /4 \right) $. Considering the coherency length of $L_\text{c}=2\pi k_0/\delta$, $a_\text{B}(z=L_\text{c})=0$ and $a_\text{A}(z=L_\text{c})= i E_0 e^{-i \phi_1}$
\begin{equation}\label{eqa:down}
E_\text{out,down-c}=i E_0 e^{-i \left(k_\text{A} z -\omega_\text{i} t \right)} e^{-i \phi_1},
\end{equation}

Equation~\eqref{eqa:up} shows that in the up-conversion photonic transition, i.e., transition from frequency $\omega_\text{i}$ to $\omega_\text{i}+\Omega$, the transmitted up-converted wave acquires the modulation phase shift of $\phi_1$. In contrast, Eq.~\eqref{eqa:down} shows that in the down-conversion photonic transition, i.e., the transition from frequency $\omega_\text{i}+\Omega$ to $\omega_\text{i}$, the wave acquires a negative phase shift, i.e., $-\phi_1$. This shows that the time-modulated medium, with the permittivity in Eq.~\ref{eqa:perm}, introduces a nonreciprocal phase shift.

\subsection{Frequency-invariant nonreciprocal radiating phase shifter.}
In the previous section, we showed that a nonreciprocal phase shift may be achieved by taking advantage of photonic transitions in a time-modulated meta-atom. However, such a nonreciprocal phase shift is accompanied with a frequency transition, which may not be always desired. Here, we propose a structure that provides a nonreciprocal phase shift without frequency alteration. Figures~\ref{Fig:coup}(b) and~\ref{Fig:coup}(c) demonstrate the operation principle of the frequency-invariant nonreciprocal radiating phase shifter formed by coupled time-modulated twin meta-atoms. Figure~\ref{Fig:coup}(b) shows a qualitative dispersion diagram of the structure for forward (green arrows) and backward (blue arrows) wave incidences, showing the general concept of the nonreciprocal phase shifting based on photonic transitions in a time-modulated meta-atom supporting two resonant frequencies, $\omega_\text{i}$ and $\omega_\text{i}+\Omega$. 

In the up-conversion, i.e., the photonic transition from $\omega_\text{i}$ to $\omega_\text{i}+\Omega$, a phase shift of $\phi$ is achieved, whereas in the down-conversion, that is, the photonic transition from $\omega_\text{i}+\Omega$ to $\omega_\text{i}$, a phase shift of $-\phi$ is introduced by the time modulation. 

The structure of the coupled time-modulated twin meta-atoms is formed by four resonators, with electric permitivitties
\begin{subequations}
	\begin{equation}
	\epsilon_1(t)=\epsilon_\text{1}+\delta_1 \cos(\Omega t+\phi_1)
	\end{equation}
	\begin{equation}
	\epsilon_1'(t)=\epsilon'_\text{1}+\delta'_1 \cos(\Omega t+\phi_1+\pi)
	\end{equation}
	\begin{equation}
	\epsilon_2(t)=\epsilon_\text{2}+\delta_2 \cos(\Omega t+\phi_2)
	\end{equation}
	\begin{equation}
	\epsilon_2'(t)=\epsilon'_\text{2}+\delta'_2 \cos(\Omega t+\phi_2+\pi)
	\end{equation}
\end{subequations}

In Fig.~\ref{Fig:coup}(c), $R_\text{r}$ and $R'_\text{r}$ represent the radiation resistances of the first and second meta-atoms, and $K$ and $K'$ denote the coupling between the arms of the first and second meta-atoms, whereas $K_\text{m}$ is the coupling between the two meta-atoms. 

In the forward incidence (left to right), the first time-modulated meta-atom, characterized with permitivitties $\epsilon_1(t)$ and $\epsilon'_1(t)$, provides a frequency-phase transition from ($\omega_\text{i}$, $0$) to ($\omega_\text{i}+\Omega$, $\phi_1$). Then, the second time-modulated meta-atom, characterized with permitivitties $\epsilon_2(t)$ and $\epsilon'_2(t)$, introduces a frequency-phase transition from ($\omega_\text{i}+\Omega$, $\phi_1$) to ($\omega_\text{i}$, $\phi_1-\phi_2$). In contrast, in the backward incidence (right to left), the second time-modulated meta-atom, characterized with permitivitties $\epsilon_2(t)$ and $\epsilon'_2(t)$, provides frequency-phase transition from ($\omega_\text{i}$, $0$) to ($\omega_\text{i}+\Omega$, $\phi_2$), and then, the first time-modulated meta-atom, characterized with permitivitties $\epsilon_1(t)$ and $\epsilon'_1(t)$, provides frequency-phase transition ($\omega_\text{i}+\Omega$, $\phi_2$) to ($\omega_\text{i}$, $\phi_2-\phi_1$). As a result, no frequency alteration occurs for both forward and backward transmitted waves, whereas a nonreciprocal phase shift is achieved, i.e., the backward transmitted wave acquires the phase shift of $\phi_2-\phi_1$ which is opposite to that of the forward transmitted wave phase $\phi_1-\phi_2$.

\section{Results}\label{sec:exper}

\subsection{Suppression of unwanted time harmonics.}

\begin{figure}[h!]
	\begin{center}
			\includegraphics[width=1\columnwidth]{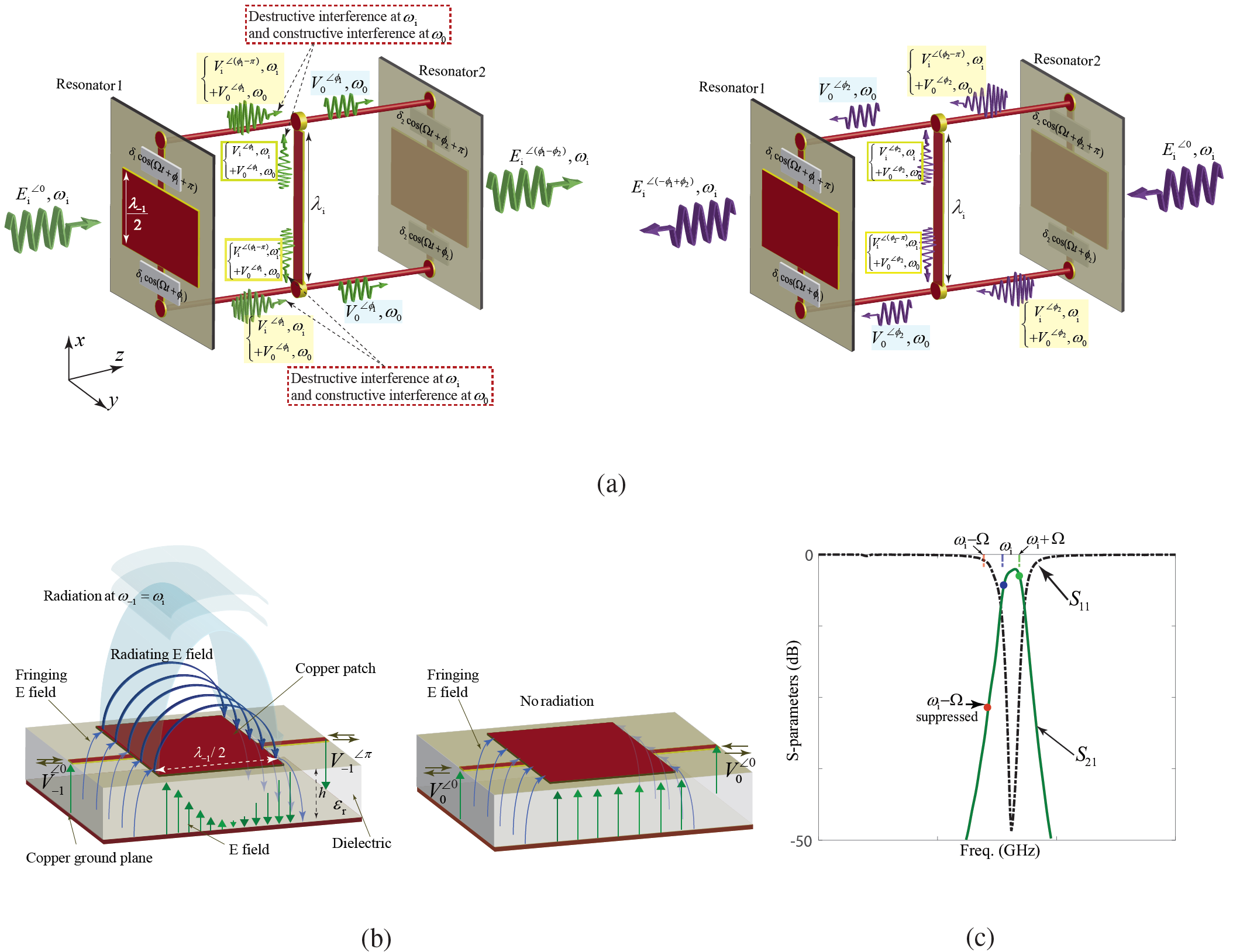}
		\caption{Architecture of the time-modulated coupled twin meta-atoms. (a) Forward and backward incidences. (b) Electric field propagation and radiation in a double-fed microstrip patch element. (c) Qualitative scattering parameters showing the suppression of $\omega_\text{-2}=\omega_\text{0}-2\omega=\omega_\text{i}-\omega$.}
		\label{Fig:coup2}
	\end{center}
\end{figure}

Figure~\ref{Fig:coup2}(a) demonstrates the structure of the coupled time-modulated twin meta-atoms, and the wave scattering and interference inside it, for forward and backward incidences. Since the time-modulated meta-atoms are periodically modulated in time, the voltage at the two arms of the coupled structure may be decomposed to Bloch-Floquet temporal harmonics, such that
\begin{subequations}\label{eq:v}
\begin{equation}
V(t)= e^{-i \omega_\text{0} t} \sum_{n=-\infty}^{\infty} V_n e^{-in ( \Omega t+\phi_1)},
\end{equation}
represents the voltage at the upper arm, and
\begin{equation}
V'(t)= e^{-i \omega_\text{0} t} \sum_{n=-\infty}^{\infty} V_n e^{-in ( \Omega t+\phi_1+\pi)}
\end{equation}
\end{subequations}
represents the voltage at the lower arm of the coupled meta-atoms. Here, $\omega_\text{-1}=\omega_\text{i}$ and $\omega_\text{0}=\omega_\text{i}+\Omega$.

It may be seen from Eq.~\eqref{eq:v} and Fig.~\ref{Fig:coup2}(a) that all odd time harmonics at the two arms of the coupled structure are $180$ degree out of phase, whereas even time harmonics are all in phase. Hence, we apply an inter-connector between the two arms of the coupled structure with the length of $\lambda_{-1}  \simeq \lambda_{0}$. Considering the $0$ and $180$ degree phase shift, respectively, between the even and odd harmonics, the middle inter-connector represents an \textit{open circuit} for even time harmonics and a \textit{short-circuit} for odd harmonics. Thereby, as shown in the top figure (forward incidence) of Fig.~\ref{Fig:coup2}(a), constructive interference occurs at all even time harmonics, i.e., $\omega_{n, \text{even}}=\omega_\text{0}+n\Omega$ with $n$ being even, and destructive interference occurs at all odd time harmonics, i.e., $\omega_{n, \text{odd}}=\omega_\text{0}+n\Omega$ with $n$ being odd. This includes suppression of the dominant odd time harmonic $\omega_\text{i}=\omega_\text{0}-\Omega$ ($n=-1$), as well as constructive interference at $\omega_\text{0}$ ($n=0$). 

Figure~\ref{Fig:coup2}(b) shows the structure of a double-fed patch radiator that is suited for efficient radiation with odd excitation. In this figure, the electric field propagation and radiation in a double-fed microstrip patch radiator is sketched, where the left sketch shows efficient radiation of the wave at $\omega_\text{i}=\omega_{-1}$ because of odd excitation ($V_{i}^{\angle 0}$ and $V_{i}^{\angle \pi}$). The right sketch in Fig.~\ref{Fig:coup2}(b) demonstrates the transmission-state operation (no radiation) of the structure at $\omega_0$ is ensured due to its even excitation ($V_{0}^{\angle 0}$ and $V_{0}^{\angle 0}$).

We shall suppress all undesired even and odd time harmonics, especially $n=-2$ which represents the lower side-band time harmonic of the incident wave. Given the flexibility of the proposed resonance structure in Fig.~\ref{Fig:coup2}(a), which is formed by two radiating patch radiators, the proposed coupled meta-atom inherently operates as a narrow bandpass filter and significantly suppresses all time harmonics that lie outside its passband. Hence, as it is shown in Fig.~\ref{Fig:coup2}(c), we design the patch radiators such that $\omega_\text{i}=\omega_\text{-1}$ and $\omega_\text{0}=\omega_\text{i}+\Omega$ lie inside the passband of the structure, whereas $\omega_\text{-2}=\omega_\text{0}-2\Omega=\omega_\text{i}-\Omega$ lies outside the passband (i.e., inside the stopband) of the structure. It should be noted that the modulation frequency $\Omega$ is an arbitrary parameter and can be adjusted to guarantee that only the desired time harmonics, here $\omega_\text{i}$ and $\omega_\text{0}$ lie inside the passband of the structure. Thus, all undesired even and odd time harmonics are suppressed and safe operation of the structure at $\omega_\text{i}$ and $\omega_\text{0}$ is guaranteed.

Therefore, only two time harmonics $\omega_\text{i}$ and $\omega_\text{0}$ will pass through the first patch element, and the signal at the upper and lower arms of the coupled structure reads
\begin{subequations}
	\begin{equation}
	V(t)=V_{-1} e^{-i ([\omega_\text{0} - \Omega]		 t+\phi_1)}+	V_{0} e^{-i \omega_\text{i} t},
	\end{equation}
	\begin{equation}
	V'(t)=V_{-1} e^{-i ([\omega_\text{0} - \Omega]		 t+\phi_1+\pi)}+	V_{0} e^{-i \omega_\text{i} t}.
	\end{equation}
\end{subequations}

Then, due to the suppression of (short-circuit) of the $-1$ time harmonic ($\omega_\text{i}$) by the middle inter-connector, the second patch radiator will be fed only at $\omega_\text{0}=\omega_\text{i}+\Omega$.

\begin{figure}[h!]
	\begin{center}
		\subfigure[]{\label{Fig:pow_flow_even}
			\includegraphics[width=0.525\columnwidth]{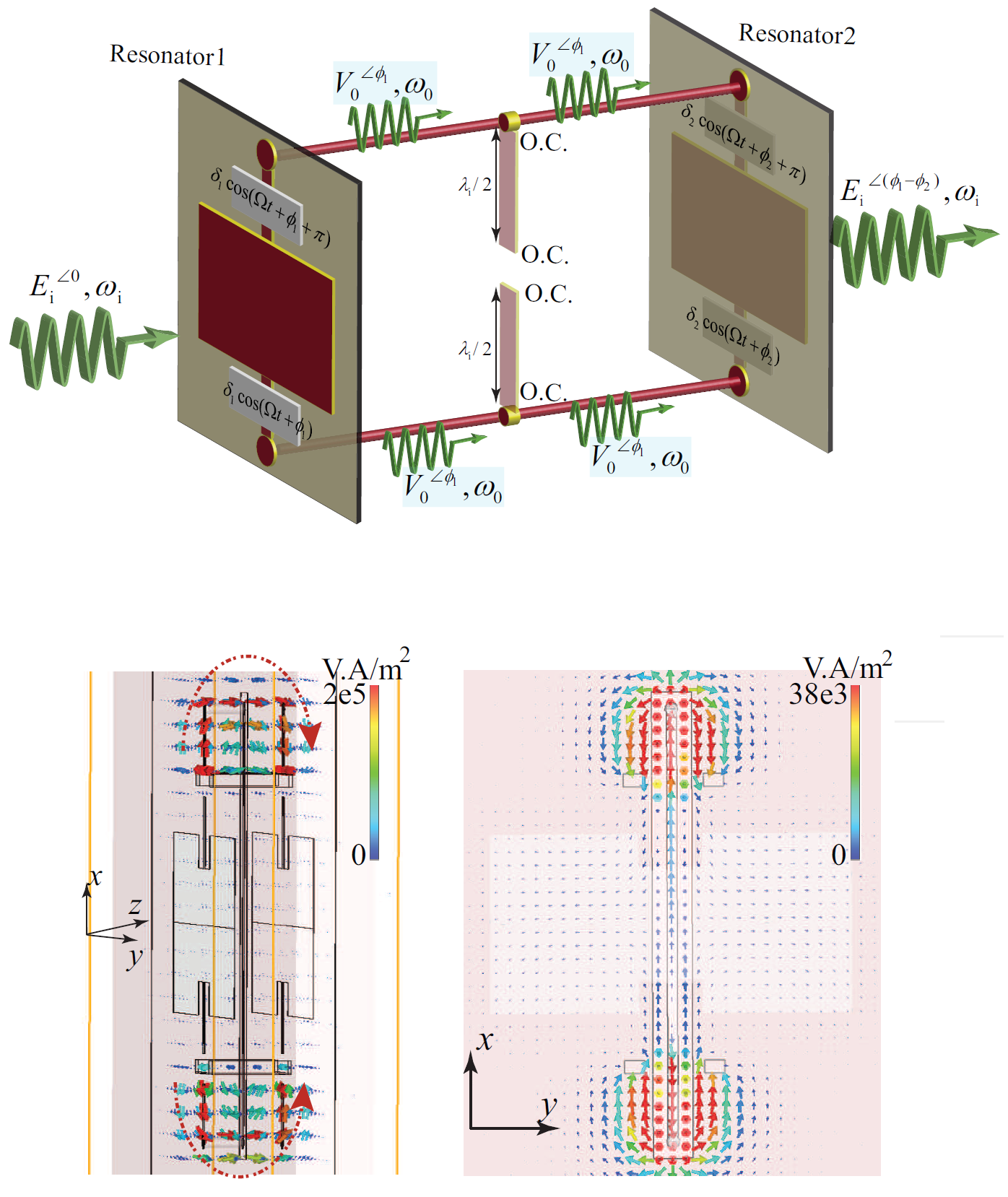}}
		\subfigure[]{\label{Fig:pow_flow_odd}
			\includegraphics[width=0.455\columnwidth]{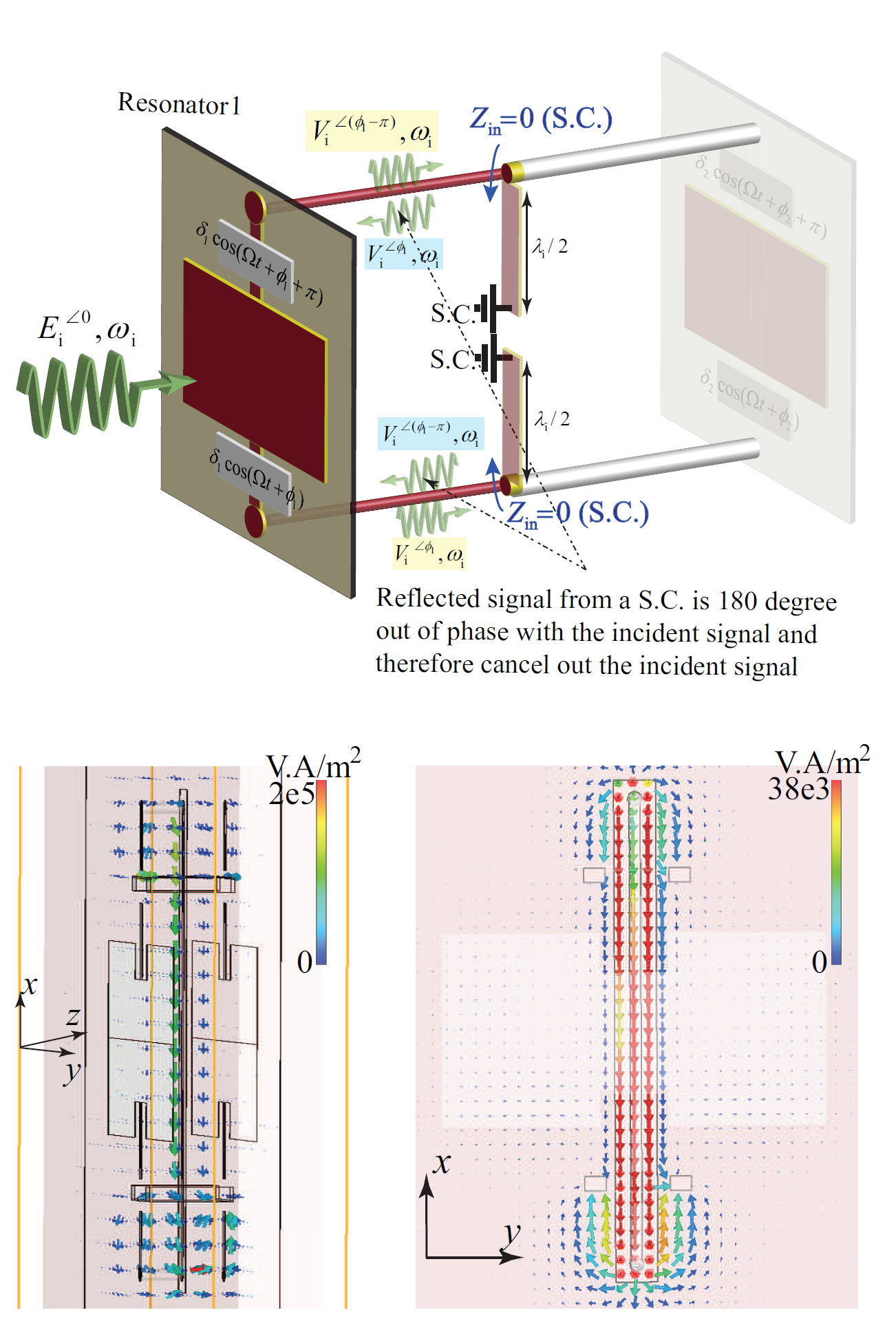}}
		\caption{Schematic representations and full-wave simulation results of the power flow inside the structure for even- and odd-mode operations of the structure at $\omega_\text{0}$ and $\omega_\text{i}$, respectively. (a) Even-mode operation at $\omega_\text{0}$, where the signals at the two horizontal interconnecting vias will reach to the second (right) meta-atom and no significant power flows through the middle inter-connector. (b) Odd-mode operation at $\omega_\text{i}$, where weak power flow exists inside the structure and the signals at the two horizontal interconnecting vias will not reach to the second (right) meta-atom, rather it flows through the middle inter-connector.}
		\label{Fig:even_odd}
	\end{center}
\end{figure}

Figures~\ref{Fig:pow_flow_even} and~\ref{Fig:pow_flow_odd} elaborate on the suppression of $\omega_\text{i}=\omega_\text{-1}$ by the middle inter-connector. Figure~\ref{Fig:pow_flow_even} shows the even-mode operation of the coupled meta-atoms at frequency $\omega_\text{0}$, where the two interconnecting horizontal vias are excited in phase, and therefore a virtual open circuit occurs at the middle of the $\lambda$ long inter-connector. This makes the middle inter-connector inactive. At the bottom of Fig.~\ref{Fig:pow_flow_even}, full-wave simulation results for the power flow for even-mode operation of the coupled meta-atoms at frequency $\omega_\text{0}$ are presented. It may be seen from this figure that strong power flows from the left meta-atom to the right meta-atom through the interconnecting via (at frequency $\omega_\text{0}$), where no significant power flows through the middle inter-connector due to the even excitation of the coupled meta-atoms ($0$ degree phase difference between the upper and lower interconnecting vias).

Figure~\ref{Fig:pow_flow_odd} shows the odd-mode operation of the coupled meta-atoms at frequency $\omega_\text{i}$, where the two interconnecting horizontal vias are excited $180$ degree out of phase, and therefore a virtual short circuit occurs at the middle of the $\lambda$ long inter-connector. This will also make a virtual short circuit at the center of two horizontal arms, hence leading to a total reflection of the incident signals with $180^\circ$ phase difference with the incident signals. As a result, the incident and reflected signals at the two horizontal interconnecting vias will cancel each other out, and therefore no signal at $\omega_\text{i}$ would exist at the two horizontal arms of the coupled meta-atoms. At the bottom of Fig.~\ref{Fig:pow_flow_odd}, full-wave simulation results for the power flow for odd-mode operation of the coupled meta-atoms at frequency $\omega_\text{i}$ is given. This figure shows that the power flow inside the coupled meta-atoms at frequency $\omega_\text{i}$ is weak in comparison to the power flow of the even-mode operation, and the power mainly flows through the middle vertical inter-connector and does not reach to the second (right) meta-atom. This is due to the odd excitation of the coupled meta-atoms at the incident frequency $\omega_\text{i}$,  where $180$ degree phase difference exists between the upper and lower horizontal interconnecting vias.

\subsection{Implementation Scenario}\label{sec:implem}
\begin{figure}
	\begin{center}
		\subfigure[]{\label{Fig:exp_unit_a}
			\includegraphics[width=0.45\columnwidth]{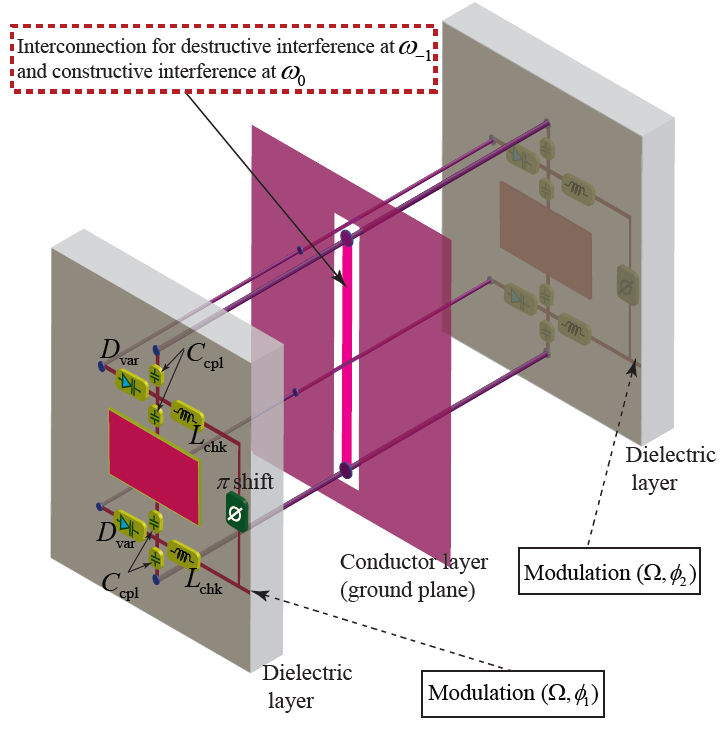}}
		\subfigure[]{\label{Fig:exp_unit_b}
			\includegraphics[width=0.55\columnwidth]{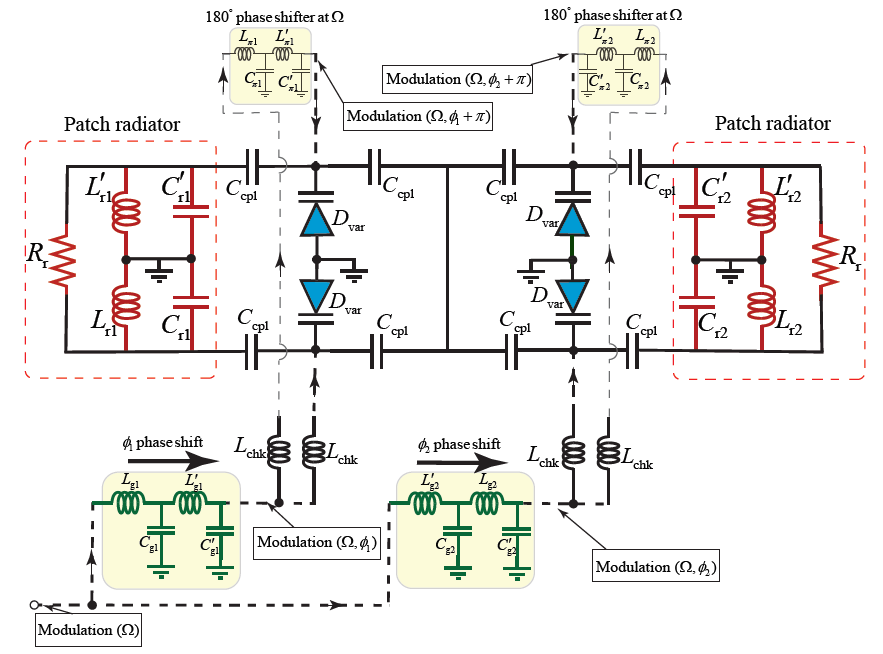}}
		\caption{Meta-atom fomed by coupled time-modulated resonators, formed by varactors and lumped elements. (a) Schematic of the structure. (b) Description of the circuit elements of the coupled twin meta-atom including the circuit model for two patch radiators, two $180^\circ$ phase shifters, two phase shifters with phases $\phi_1$ and $\phi_2$, respectively, four varactor diodes, four choke inductors, and eight decoupling capacitances.}
		\label{Fig:exp_unit}
	\end{center}
\end{figure}

Figure~\ref{Fig:exp_unit_a} shows different elements of the coupled time-modulated twin meta-atoms, providing three major operations, that is, reception of the incoming wave from one side, application of a nonreciprocal phase shift to the wave, and re-radiation of the processed phase-shifted wave to the other side of the structure. The meta-atom in Fig.~\ref{Fig:exp_unit_a} is formed by coupled time-modulated resonators, constituted by varactors and lumped elements. Figure~\ref{Fig:exp_unit_b} describes the circuit elements of the coupled twin meta-atom including the circuit model for the two patch radiators, two $180^\circ$ phase shifters, two phase shifters with phases $\phi_1$ and $\phi_2$, respectively, the four varactor diodes $D_\text{var}$, four choke inductors $L_\text{chk}$, and eight decoupling capacitances $C_\text{cpl}$. Two inductances $L_\text{chk}$ and four capacitances $C_\text{cpl}$ are employed at each side of the coupled meta-atoms to efficiently prevent the leakage of the incident wave to the modulation path, as well as decouple the two modulation signals (with $180^\circ$ phase difference) at the upper and lower side of the meta-atom.
	
\begin{figure}
	\begin{center}
		\subfigure[]{\label{Fig:exp_comp}
			\includegraphics[width=0.5\columnwidth]{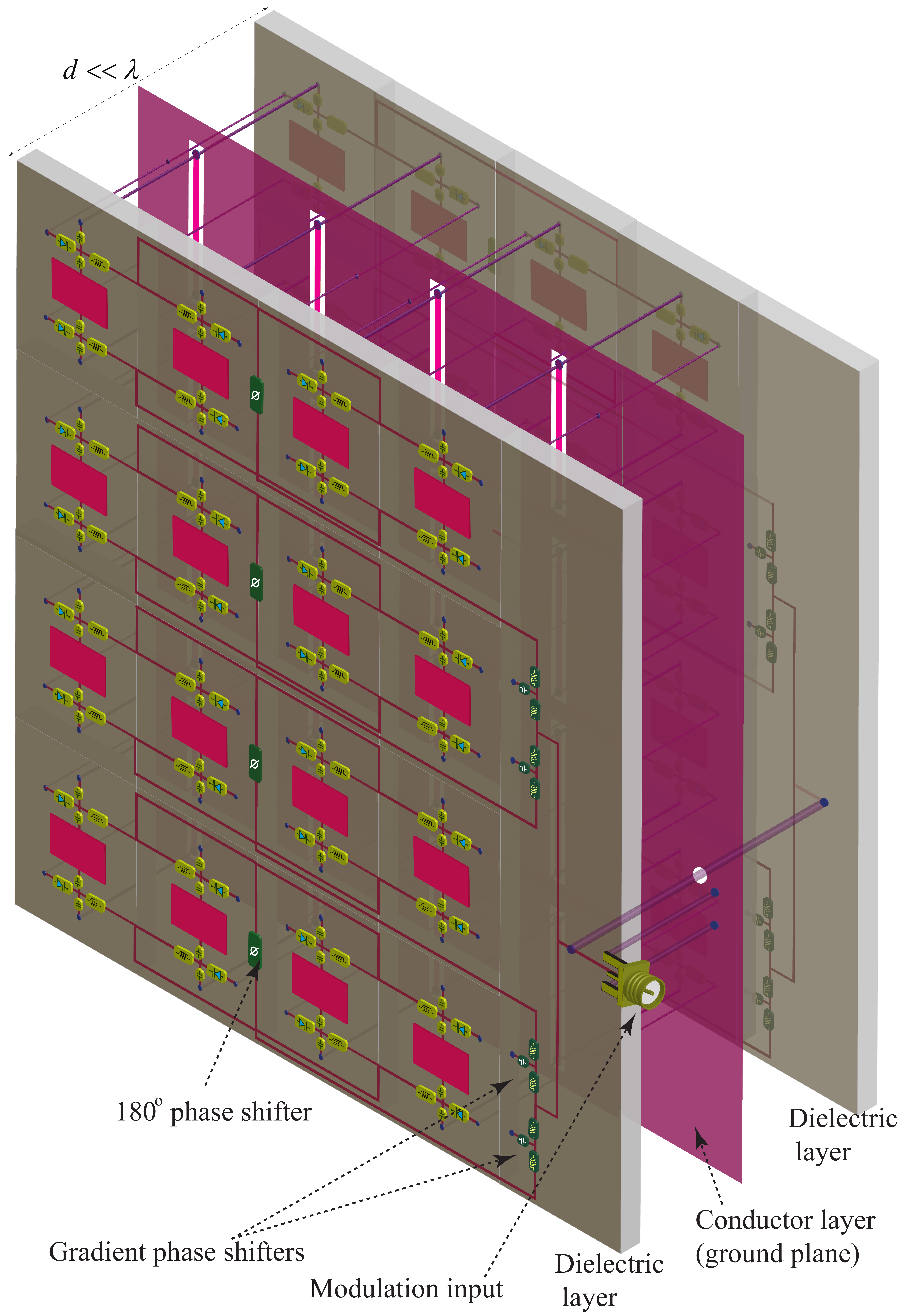} }\\
		\subfigure[]{\label{Fig:photo_a}
			\includegraphics[width=0.44\columnwidth]{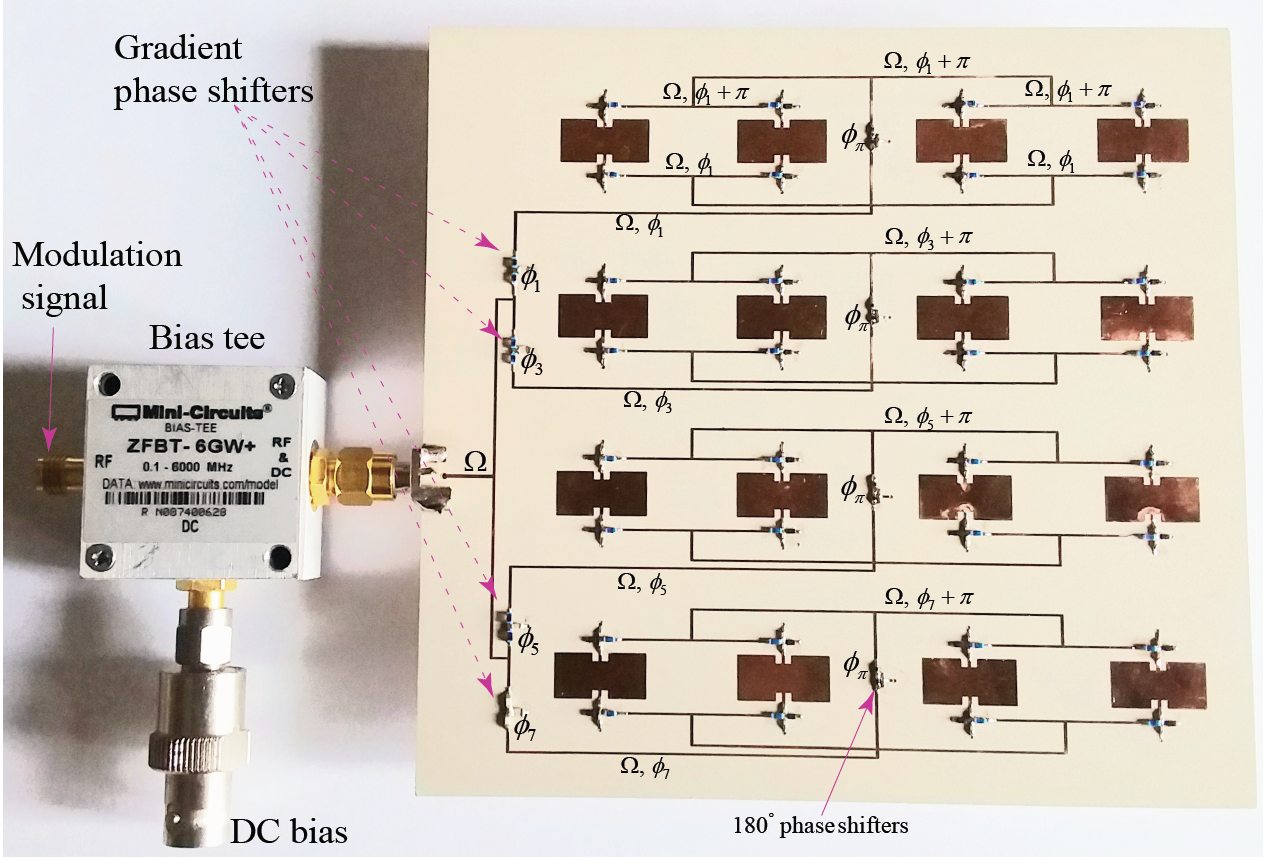} }
		\subfigure[]{\label{Fig:photo_b}
			\includegraphics[width=0.42\columnwidth]{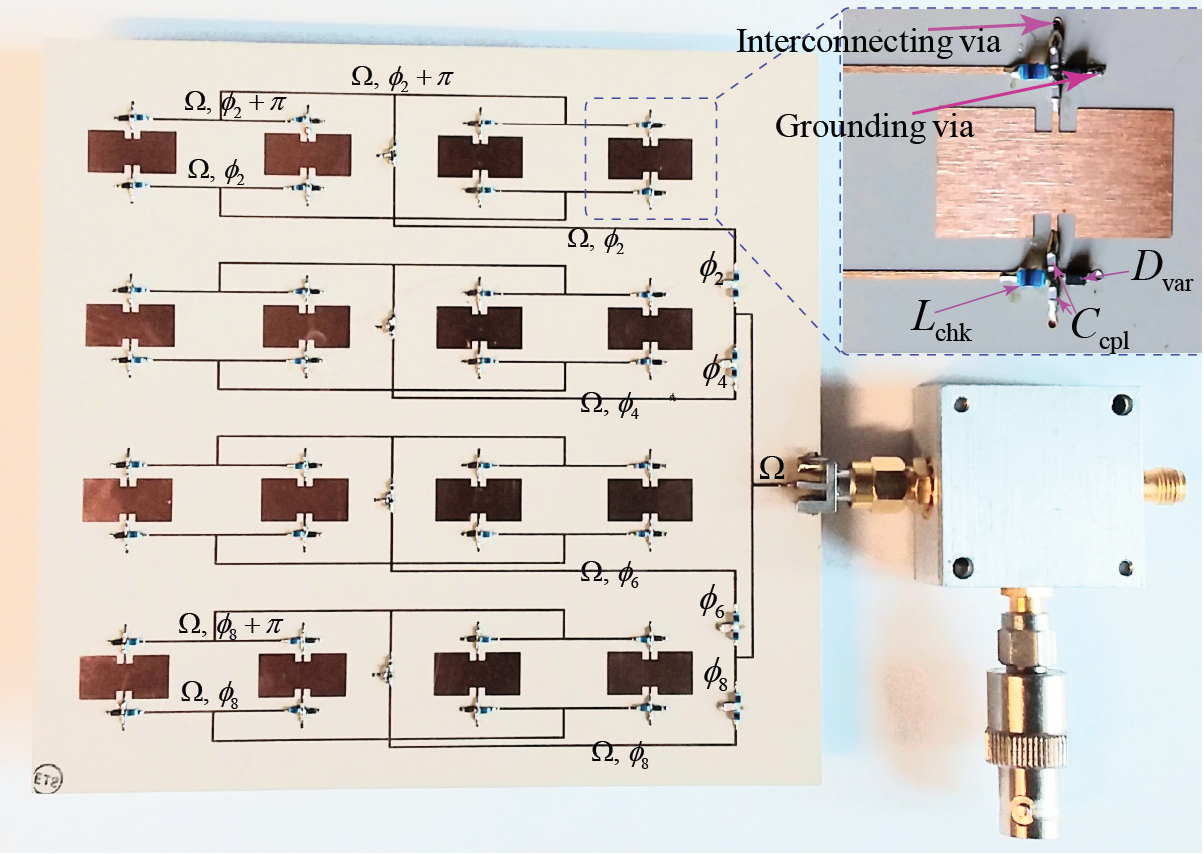} }	
		\caption{Experimental demonstration of full-duplex nonreciprocal-beam-steering transmissive phase-gradient metasurface. (a) Architecture of the implemented metasurface composed of $4 \times4$ meta-atoms. (b) and (c) Top and bottom views of the fabricated prototype, respectively.}
		\label{Fig:photo}
	\end{center}
\end{figure}

Figure~\ref{Fig:exp_comp} depicts a schematic of the complete phase-gradient metasurface comprising $4 \times 4$ coupled twin time-modulated meta-atoms, shown in Fig.~\ref{Fig:exp_unit_a}. The modulation signal is fed to the metasurface via a SMA connector. The metasurface includes eight gradient phase shifters ($\phi_1$ to $\phi_8$), i.e. four phase shifters in each side, providing the required modulation phase shifts for nonreciprocal beam steering. In addition, eight $180^\circ$ phase shifters ($\phi_\pi$) are utilized for achieving the $\pi$ phase-shifted version of each gradient phase shifted modulation signal. 

\subsection{Experimental Demonstration}\label{sec:exp}
We next experimentally demonstrate the full-duplex nonreciprocal-beam-steering time-modulated transmissive metasurface. Figure~\ref{Fig:photo_a} shows the top view, which includes gradient phase shifters $\phi_1$, $\phi_3$, $\phi_5$ and $\phi_7$, and Figure~\ref{Fig:photo_b} shows the bottom view, which includes gradient phase shifters $\phi_2$, $\phi_4$, $\phi_6$ and $\phi_8$. A DC signal is applied to varactors to achieve the desired average capacitance (average permittivity). A bias-tee is used to separate the DC bias and the modulation signal. We employ a total number of 64 SMV1247-079LF varactors ($D_\text{var}$) from Skyworks Solutions Inc., 64 inductors of $L_\text{chk}=20$~nH, and 128 decoupling capacitances of $C_\text{cpl}=5$~pF. The metasurface is fabricated as a three-layer circuit, i.e., three conductor layers and two dielectric layers, made of
Rogers RO3210 with 50 mils height ($d=100$ mils). The incident wave and modulation parameters are set as $\omega_\text{i}=5.28$~GHz, $\Omega=50$~MHz, and $\omega_\text{0}=\omega_\text{i}+\Omega=5.33$~GHz.

\begin{figure}
	\begin{center}	
			\includegraphics[width=1\columnwidth]{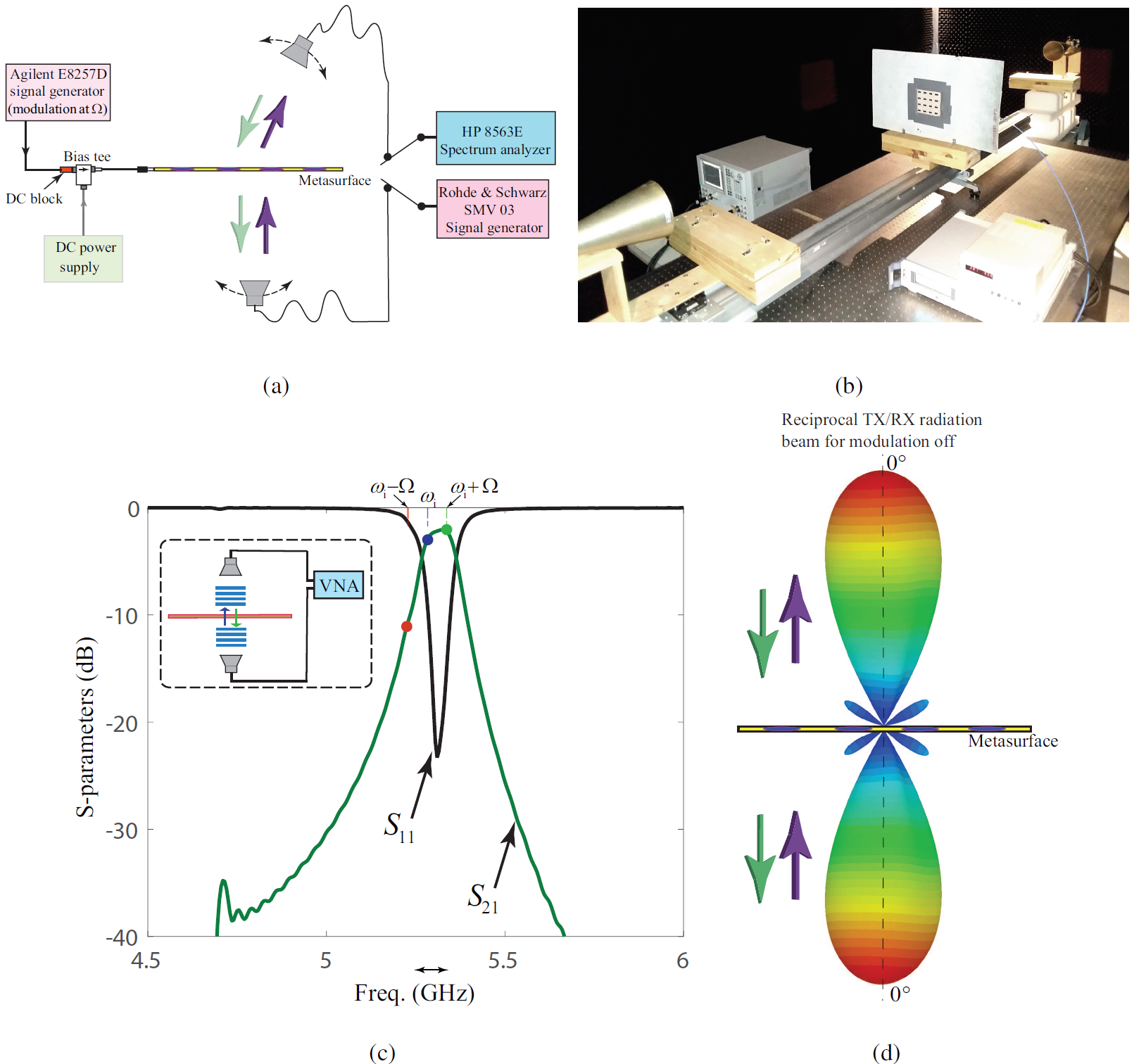} 
		\caption{Experimental demonstration of nonreciprocal radiation beam metasurface. (a) Schematic of the measurement set-up consisting of two signal generators, a spectrum analyzer and a DC power supply. (b) A photo of the measurement set-up in an anechoic chamber. (c) Experimental results for the scattering parameters of the metasurface measured using a vector network analyzer. (d) Simulation and experimental results for the reciprocal radiation pattern of the un-modulated metasurface.}
		\label{Fig:meas}
	\end{center}
\end{figure}

Figure~\ref{Fig:meas}(a) depicts a schematic of the measurement set-up consisting of two signal generators, a spectrum analyzer and a DC power supply. A photo of the measurement set-up is shown in Fig.~\ref{Fig:meas}(b). Figure~\ref{Fig:meas}(c) plots the scattering parameters of the fabricated nonreciprocal radiation beam metasurface. This figure highlights three different frequencies, $\omega_\text{i}-\Omega$, $\omega_\text{i}$ and $\omega_\text{i}+\Omega$. As described in the previous section, two time harmonics $\omega_\text{i}$ and $\omega_\text{i}+\Omega$ lie inside the passband of the structure, whereas  $\omega_\text{i}-\Omega$ lies in the stopband of the structure and is being suppressed. The scattering parameters are measured using an $E8361C$ Agilent vector network analyzer. For the sake of comparison, we first measured the radiation beam of the un-modulated metasurface, i.e., $\Omega=0$. Figure~\ref{Fig:meas}(d) plots the radiation pattern of the un-modulated metasurface, which is reciprocal, that is, provides identical radiation beams for both transmission and reception states.

\begin{figure}
	\begin{center}
			\includegraphics[width=1\columnwidth]{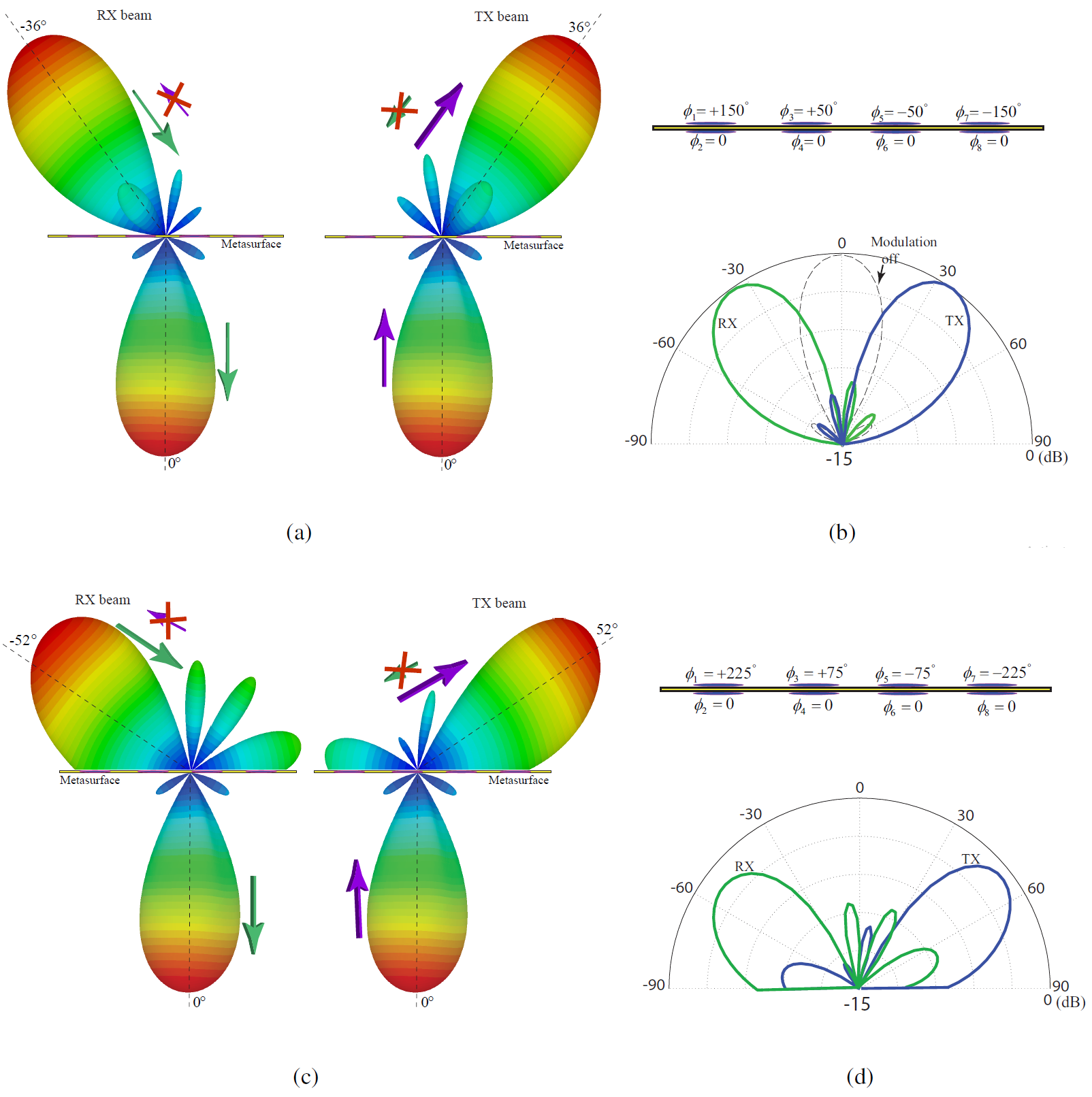} 
		\caption{Full-wave simulation results for angle-symmetric transmission and reception radiation patterns of the nonreciprocal-beam-metasurfcae. (a) and (b) Simulation results for the normalized radiation beam at the two sides of the metasurface for reception and transmission for $\theta_\text{TX}=-\theta_\text{RX}=36^\circ$. (c) and (d) Simulation results for the normalized radiation beam at the two sides of the metasurface for reception and transmission for $\theta_\text{TX}=-\theta_\text{RX}=52^\circ$.}
		\label{Fig:rad_patt}
	\end{center}
\end{figure}

Figures~\ref{Fig:rad_patt}(a) and~\ref{Fig:rad_patt}(b) plot the full-wave simulation results for the nonreciprocal \textit{angle-symmetric} transmission and reception radiation patterns of the nonreciprocal radiation beam metasurface for $\theta_\text{1,TX}=0$ and $\theta_\text{1,RX}=0^\circ$, $\theta_\text{TX}=-\theta_\text{RX}=36^\circ$. Figures~\ref{Fig:rad_patt}(c) and~\ref{Fig:rad_patt}(d) plot the full-wave simulation results for the nonreciprocal \textit{angle-symmetric} transmission and reception radiation patterns of the nonreciprocal radiation beam metasurface for $\theta_\text{1,TX}=0$ and $\theta_\text{1,RX}=0^\circ$, $\theta_\text{TX}=-\theta_\text{RX}=52^\circ$. The relation between the angles of reception and transmission and the required nonreciprocal gradient phase shifter for each time-modulated meta-atom is governed by Eqs.~\eqref{eq:phi_tx},~\eqref{eq:phi_rx},~\eqref{eq:tx}, and~\eqref{eq:rx}.

\begin{figure}
	\begin{center}
			\includegraphics[width=0.9\columnwidth]{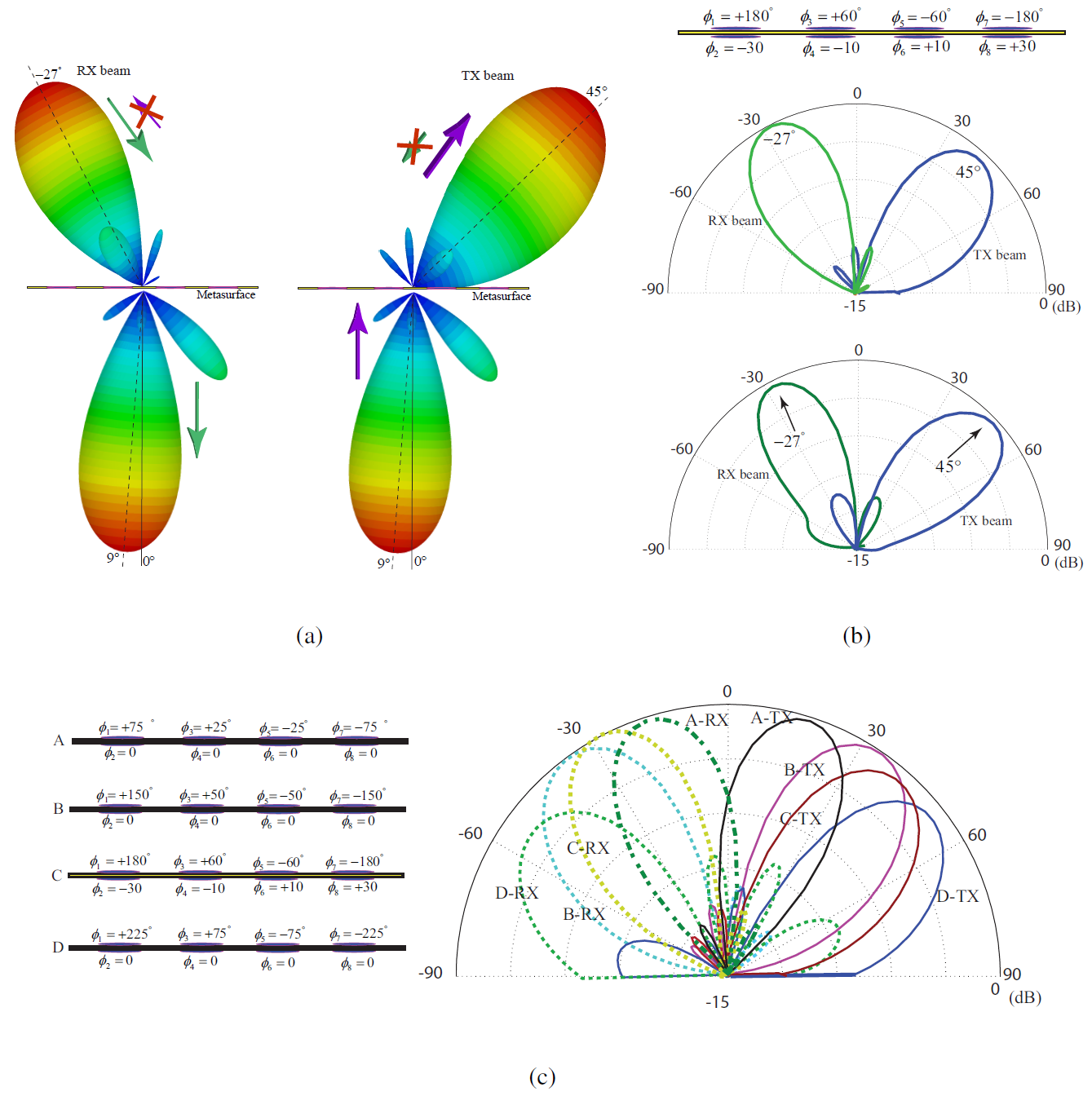} 
		\caption{Full-wave simulation and experimental results. (a) Angle-asymmetric transmission and reception radiation patterns of the nonreciprocal-beam-metasurface. Simulation results show the normalized radiation beam at the two sides of the metasurface for (left) reception and (right) transmission, corresponding to $\theta_\text{TX,2}=45^\circ$, $\theta_\text{RX,2}=-27^\circ$, $\theta_\text{TX,1}=\theta_\text{RX,1}=9^\circ$. (b) Results for the TX and RX normalized radiation beams in comparison with reciprocal normalized radiation beam for modulation off. (top) Simulation, and (bottom) experimental results. (c) Full-duplex angle-symmetric/asymmetric beam steering.}
		\label{Fig:rad_patt_as}
	\end{center}
\end{figure}

Figures~\ref{Fig:rad_patt_as}(a) and (b) plot the full-wave simulation and experimental results for the nonreciprocal \textit{angle-asymmetric} transmission and reception radiation patterns of the nonreciprocal radiation beam metasurface for $\theta_\text{1,TX}=\theta_\text{1,RX}=9^\circ$, $\theta_\text{2,TX}=45^\circ$ and $\theta_\text{2,RX}=-27^\circ$. In this particular case, the experimental isolation
between the transmission and reception radiation patters at specified transmission radiation angle ($\theta_\text{2,TX}=45^\circ$) is about $15.8$ dB, and the isolation at specified reception radiation angle ($\theta_\text{2,RX}=-27^\circ$) is about $10.4$ dB. To achieve higher isolation levels, one may change the modulation parameters or use a more directive metasurface by increasing the number of coupled twin meta-atoms. Figure~\ref{Fig:rad_patt_as}(c) plots the full-wave simulation results demonstrating the full-duplex angle-symmetric/asymmetric beam steering functionality of the time-modulated gradient transmissive metasurface. In this figure, four different gradient profiles are considered, that is, A, B, C and D.

\subsection{Discussion}\label{sec:disc}

In conclusion, we have presented a theoretical and experimental strategy for the realization of full-duplex nonreciprocal-beam-steering transmissive metasurfaces. Such a strategy takes advantage of the unique and peculiar properties of asymmetric frequency-phase transitions in coupled time-modulated twin meta-atoms. The proposed coupled meta-atom architecture leverages the extraordinary operation of periodic double-fed time-modulated patch antennas, and their associated peculiar frequency-phase transitions. The metasurfaces are composed of an array of coupled time-modulated meta-atoms, each of which is fed by a specified modulation phase, such that a phase-gradient modulation is applied to coupled meta-atoms. As a result of such gradient modulation phase, a nonreciprocal-phase-gradient metasurface is realized and therefore, the incoming (received) and outgoing (transmitted) waves are radiated at different radiation angles. We showed that the proposed coupled time-modulated meta-atoms inherently prohibit the excitation of undesired time harmonics. This results in a high conversion efficiency, i.e., less than $1$ dB insertion loss, which is of paramount importance for practical applications such as for instance point to point full-duplex communications. It should be noted that the low insertion loss of the metasurface is supported by the injection of the energy from the time modulation to the incident wave. The proposed metasurface is expected to find applications in point to point full-duplex wireless communication systems that require highly directive radiation beams, with possibility of different half-power beamwidths, for the transmission and reception states. This includes satellite communications and cellular communication systems. 


\section{Conclusion}\label{sec:conc}

%


\bibliographystyle{naturemag}
\bibliography{Taravati_Reference}

\begin{thebibliography}{10}
\expandafter\ifx\csname url\endcsname\relax
  \def\url#1{\texttt{#1}}\fi
\expandafter\ifx\csname urlprefix\endcsname\relax\def\urlprefix{URL }\fi
\providecommand{\bibinfo}[2]{#2}
\providecommand{\eprint}[2][]{\url{#2}}

\bibitem{Alu_PRB_2015}
\bibinfo{author}{Hadad, Y.}, \bibinfo{author}{Sounas, D.~L.} \&
  \bibinfo{author}{Al\`{u}, A.}
\newblock \bibinfo{title}{Space-time gradient metasurfaces}.
\newblock \emph{\bibinfo{journal}{{Phys. Rev. B}}}
  \textbf{\bibinfo{volume}{92}}, \bibinfo{pages}{100304}
  (\bibinfo{year}{2015}).

\bibitem{Taravati_2016_NR_Nongyro}
\bibinfo{author}{Taravati, S.}, \bibinfo{author}{Khan, B.~A.},
  \bibinfo{author}{Gupta, S.}, \bibinfo{author}{Achouri, K.} \&
  \bibinfo{author}{Caloz, C.}
\newblock \bibinfo{title}{Nonreciprocal nongyrotropic magnetless metasurface}.
\newblock \emph{\bibinfo{journal}{{IEEE Trans. Antennas Propagat.}}}
  \textbf{\bibinfo{volume}{65}}, \bibinfo{pages}{3589--3597}
  (\bibinfo{year}{2017}).

\bibitem{Fan_APL_2016}
\bibinfo{author}{Shi, Y.} \& \bibinfo{author}{Fan, S.}
\newblock \bibinfo{title}{Dynamic non-reciprocal meta-surfaces with arbitrary
  phase reconfigurability based on photonic transition in meta-atoms}.
\newblock \emph{\bibinfo{journal}{{Appl. Phys. Lett.}}}
  \textbf{\bibinfo{volume}{108}}, \bibinfo{pages}{021110}
  (\bibinfo{year}{2016}).

\bibitem{wang2018extreme}
\bibinfo{author}{Wang, X.} \emph{et~al.}
\newblock \bibinfo{title}{Extreme asymmetry in metasurfaces via evanescent
  fields engineering: Angular-asymmetric absorption}.
\newblock \emph{\bibinfo{journal}{{Phys. Rev. Lett.}}}
  \textbf{\bibinfo{volume}{121}}, \bibinfo{pages}{256802}
  (\bibinfo{year}{2018}).

\bibitem{zhang2019breaking}
\bibinfo{author}{Zhang, L.} \emph{et~al.}
\newblock \bibinfo{title}{Breaking reciprocity with space-time-coding digital
  metasurfaces}.
\newblock \emph{\bibinfo{journal}{{Adv. Mater.}}} \bibinfo{pages}{1904069}.

\bibitem{Grbic2019serrodyne}
\bibinfo{author}{Wu, Z.} \& \bibinfo{author}{Grbic, A.}
\newblock \bibinfo{title}{Serrodyne frequency translation using time-modulated
  metasurfaces}.
\newblock \emph{\bibinfo{journal}{arXiv preprint arXiv:1905.06792}}
  (\bibinfo{year}{2019}).

\bibitem{sounas2017non}
\bibinfo{author}{Sounas, D.~L.} \& \bibinfo{author}{Al{\`u}, A.}
\newblock \bibinfo{title}{Non-reciprocal photonics based on time modulation}.
\newblock \emph{\bibinfo{journal}{{Nat. Photonics}}}
  \textbf{\bibinfo{volume}{11}}, \bibinfo{pages}{774} (\bibinfo{year}{2017}).

\bibitem{zang2019nonreciprocal}
\bibinfo{author}{Zang, J.}, \bibinfo{author}{Wang, X.},
  \bibinfo{author}{Alvarez-Melcon, A.} \& \bibinfo{author}{Gomez-Diaz, J.}
\newblock \bibinfo{title}{Nonreciprocal yagi-uda filtering antennas}.
\newblock \emph{\bibinfo{journal}{arXiv preprint arXiv:1906.06418}}
  (\bibinfo{year}{2019}).

\bibitem{Taravati_Kishk_TAP_2019}
\bibinfo{author}{Taravati, S.} \& \bibinfo{author}{Kishk, A.~A.}
\newblock \bibinfo{title}{Advanced wave engineering via obliquely illuminated
  space-time-modulated slab}.
\newblock \emph{\bibinfo{journal}{{IEEE Trans. Antennas Propagat.}}}
  \textbf{\bibinfo{volume}{67}}, \bibinfo{pages}{270--281}
  (\bibinfo{year}{2019}).

\bibitem{ptitcyn2019time}
\bibinfo{author}{Ptitcyn, G.}, \bibinfo{author}{Mirmoosa, M.} \&
  \bibinfo{author}{Tretyakov, S.}
\newblock \bibinfo{title}{Time-modulated meta-atoms}.
\newblock \emph{\bibinfo{journal}{{Phys. Rev. Res.}}}
  \textbf{\bibinfo{volume}{1}}, \bibinfo{pages}{023014} (\bibinfo{year}{2019}).

\bibitem{cassedy1965waves}
\bibinfo{author}{Cassedy, E.~S.}
\newblock \bibinfo{title}{Waves guided by a boundary with time-space periodic
  modulation}.
\newblock In \emph{\bibinfo{booktitle}{Proceedings of the Institution of
  Electrical Engineers}}, vol. \bibinfo{volume}{112}, \bibinfo{pages}{269--279}
  (\bibinfo{organization}{IET}, \bibinfo{year}{1965}).

\bibitem{Cassedy_PIEEE_1967}
\bibinfo{author}{Cassedy, E.~S.}
\newblock \bibinfo{title}{Dispersion relations in time-space periodic media:
  part \textsc{II}-unstable interactions}.
\newblock \emph{\bibinfo{journal}{{Proc. IEEE}}} \textbf{\bibinfo{volume}{55}},
  \bibinfo{pages}{1154 -- 1168} (\bibinfo{year}{1967}).

\bibitem{Taravati_PRAp_2018}
\bibinfo{author}{Taravati, S.}
\newblock \bibinfo{title}{Giant linear nonreciprocity, zero reflection, and
  zero band gap in equilibrated space-time-varying media}.
\newblock \emph{\bibinfo{journal}{Phys. Rev. Appl.}}
  \textbf{\bibinfo{volume}{9}}, \bibinfo{pages}{064012} (\bibinfo{year}{2018}).

\bibitem{inampudi2019rigorous}
\bibinfo{author}{Inampudi, S.}, \bibinfo{author}{Salary, M.~M.},
  \bibinfo{author}{Jafar-Zanjani, S.} \& \bibinfo{author}{Mosallaei, H.}
\newblock \bibinfo{title}{Rigorous space-time coupled-wave analysis for
  patterned surfaces with temporal permittivity modulation}.
\newblock \emph{\bibinfo{journal}{{Opt. Mater. Express}}}
  \textbf{\bibinfo{volume}{9}}, \bibinfo{pages}{162--182}
  (\bibinfo{year}{2019}).

\bibitem{elnaggar2019generalized}
\bibinfo{author}{Elnaggar, S.~Y.} \& \bibinfo{author}{Milford, G.~N.}
\newblock \bibinfo{title}{Generalized space-time periodic circuits for
  arbitrary structures}.
\newblock \emph{\bibinfo{journal}{arXiv preprint arXiv:1901.08698}}
  (\bibinfo{year}{2019}).

\bibitem{Taravati_Kishk_PRB_2018}
\bibinfo{author}{Taravati, S.} \& \bibinfo{author}{Kishk, A.~A.}
\newblock \bibinfo{title}{Dynamic modulation yields one-way beam splitting}.
\newblock \emph{\bibinfo{journal}{{Phys. Rev. B}}}
  \textbf{\bibinfo{volume}{99}}, \bibinfo{pages}{075101}
  (\bibinfo{year}{2019}).

\bibitem{wang2018photonic}
\bibinfo{author}{Wang, N.}, \bibinfo{author}{Zhang, Z.-Q.} \&
  \bibinfo{author}{Chan, C.}
\newblock \bibinfo{title}{Photonic floquet media with a complex time-periodic
  permittivity}.
\newblock \emph{\bibinfo{journal}{{Phys. Rev. B}}}
  \textbf{\bibinfo{volume}{98}}, \bibinfo{pages}{085142}
  (\bibinfo{year}{2018}).

\bibitem{Taravati_Kishk_MicMag_2019}
\bibinfo{author}{Taravati, S.} \& \bibinfo{author}{Kishk, A.~A.}
\newblock \bibinfo{title}{Space-time modulation: Principles and applications}.
\newblock \emph{\bibinfo{journal}{IEEE Microwave Magazine, arXiv preprint
  arXiv:1903.01272}}  (\bibinfo{year}{2019}).

\bibitem{zurita2009reflection}
\bibinfo{author}{Zurita-S{\'a}nchez, J.~R.}, \bibinfo{author}{Halevi, P.} \&
  \bibinfo{author}{Cervantes-Gonzalez, J.~C.}
\newblock \bibinfo{title}{Reflection and transmission of a wave incident on a
  slab with a time-periodic dielectric function}.
\newblock \emph{\bibinfo{journal}{{Phys. Rev. A}}}
  \textbf{\bibinfo{volume}{79}}, \bibinfo{pages}{053821}
  (\bibinfo{year}{2009}).

\bibitem{martinez2018parametric}
\bibinfo{author}{Mart{\'\i}nez-Romero, J.~S.} \& \bibinfo{author}{Halevi, P.}
\newblock \bibinfo{title}{Parametric resonances in a temporal photonic crystal
  slab}.
\newblock \emph{\bibinfo{journal}{{Phys. Rev. A}}}
  \textbf{\bibinfo{volume}{98}}, \bibinfo{pages}{053852}
  (\bibinfo{year}{2018}).

\bibitem{salary2018time}
\bibinfo{author}{Salary, M.~M.}, \bibinfo{author}{Jafar-Zanjani, S.} \&
  \bibinfo{author}{Mosallaei, H.}
\newblock \bibinfo{title}{Time-varying metamaterials based on graphene-wrapped
  microwires: Modeling and potential applications}.
\newblock \emph{\bibinfo{journal}{{Phys. Rev. B}}}
  \textbf{\bibinfo{volume}{97}}, \bibinfo{pages}{115421}
  (\bibinfo{year}{2018}).

\bibitem{mirmoosa2019time}
\bibinfo{author}{Mirmoosa, M.}, \bibinfo{author}{Ptitcyn, G.},
  \bibinfo{author}{Asadchy, V.} \& \bibinfo{author}{Tretyakov, S.}
\newblock \bibinfo{title}{Time-varying reactive elements for extreme
  accumulation of electromagnetic energy}.
\newblock \emph{\bibinfo{journal}{{Phys. Rev. Appl.}}}
  \textbf{\bibinfo{volume}{11}}, \bibinfo{pages}{014024}
  (\bibinfo{year}{2019}).

\bibitem{Fan_NPH_2009}
\bibinfo{author}{Yu, Z.} \& \bibinfo{author}{Fan, S.}
\newblock \bibinfo{title}{Complete optical isolation created by indirect
  interband photonic transitions}.
\newblock \emph{\bibinfo{journal}{{Nat. Photonics}}}
  \textbf{\bibinfo{volume}{3}}, \bibinfo{pages}{91 -- 94}
  (\bibinfo{year}{2009}).

\bibitem{li2019nonreciprocal}
\bibinfo{author}{Li, J.}, \bibinfo{author}{Shen, C.}, \bibinfo{author}{Zhu,
  X.}, \bibinfo{author}{Xie, Y.} \& \bibinfo{author}{Cummer, S.~A.}
\newblock \bibinfo{title}{Nonreciprocal sound propagation in space-time
  modulated media}.
\newblock \emph{\bibinfo{journal}{{Phys. Rev. B}}}
  \textbf{\bibinfo{volume}{99}}, \bibinfo{pages}{144311}
  (\bibinfo{year}{2019}).

\bibitem{oudich2019space}
\bibinfo{author}{Oudich, M.}, \bibinfo{author}{Deng, Y.}, \bibinfo{author}{Tao,
  M.} \& \bibinfo{author}{Jing, Y.}
\newblock \bibinfo{title}{Space-time phononic crystals with anomalous
  topological edge states}.
\newblock \emph{\bibinfo{journal}{arXiv preprint arXiv:1904.02711}}
  (\bibinfo{year}{2019}).

\bibitem{correas2018magnetic}
\bibinfo{author}{Correas-Serrano, D.}, \bibinfo{author}{Al{\`u}, A.} \&
  \bibinfo{author}{Gomez-Diaz, J.}
\newblock \bibinfo{title}{Magnetic-free nonreciprocal photonic platform based
  on time-modulated graphene capacitors}.
\newblock \emph{\bibinfo{journal}{{Phys. Rev. B}}}
  \textbf{\bibinfo{volume}{98}}, \bibinfo{pages}{165428}
  (\bibinfo{year}{2018}).

\bibitem{liu2018huygens}
\bibinfo{author}{Liu, M.}, \bibinfo{author}{Powell, D.~A.},
  \bibinfo{author}{Zarate, Y.} \& \bibinfo{author}{Shadrivov, I.~V.}
\newblock \bibinfo{title}{Huygens’ metadevices for parametric waves}
  \textbf{\bibinfo{volume}{8}}, \bibinfo{pages}{031077} (\bibinfo{year}{2018}).

\bibitem{du2019simulation}
\bibinfo{author}{Du, Z.-X.}, \bibinfo{author}{Li, A.}, \bibinfo{author}{Zhang,
  X.~Y.} \& \bibinfo{author}{Sievenpiper, D.~F.}
\newblock \bibinfo{title}{A simulation technique for radiation properties of
  time-varying media based on frequency-domain solvers}.
\newblock \emph{\bibinfo{journal}{IEEE Access}} \textbf{\bibinfo{volume}{7}},
  \bibinfo{pages}{112375--112383} (\bibinfo{year}{2019}).

\bibitem{wentz1966nonreciprocal}
\bibinfo{author}{Wentz, J.}
\newblock \bibinfo{title}{A nonreciprocal electrooptic device}.
\newblock \emph{\bibinfo{journal}{Proc. IEEE}} \textbf{\bibinfo{volume}{54}},
  \bibinfo{pages}{97--98} (\bibinfo{year}{1966}).

\bibitem{Fan_PRL_109_2012}
\bibinfo{author}{Lira, H.}, \bibinfo{author}{Yu, Z.}, \bibinfo{author}{Fan, S.}
  \& \bibinfo{author}{Lipson, M.}
\newblock \bibinfo{title}{Electrically driven nonreciprocity induced by
  interband photonic transition on a silicon chip}.
\newblock \emph{\bibinfo{journal}{{Phys. Rev. Lett.}}}
  \textbf{\bibinfo{volume}{109}}, \bibinfo{pages}{033901}
  (\bibinfo{year}{2012}).

\bibitem{Wang_TMTT_2014}
\bibinfo{author}{Qin, S.}, \bibinfo{author}{Xu, Q.} \& \bibinfo{author}{Wang,
  Y.~E.}
\newblock \bibinfo{title}{Nonreciprocal components with distributedly modulated
  capacitors}.
\newblock \emph{\bibinfo{journal}{{IEEE Trans. Microw. Theory Techn.}}}
  \textbf{\bibinfo{volume}{62}}, \bibinfo{pages}{2260--2272}
  (\bibinfo{year}{2014}).

\bibitem{zanjani2014one}
\bibinfo{author}{Zanjani, M.~B.}, \bibinfo{author}{Davoyan, A.~R.},
  \bibinfo{author}{Mahmoud, A.~M.}, \bibinfo{author}{Engheta, N.} \&
  \bibinfo{author}{Lukes, J.~R.}
\newblock \bibinfo{title}{One-way phonon isolation in acoustic waveguides}.
\newblock \emph{\bibinfo{journal}{{Appl. Phys. Lett.}}}
  \textbf{\bibinfo{volume}{104}}, \bibinfo{pages}{081905}
  (\bibinfo{year}{2014}).

\bibitem{Bahl_2015non}
\bibinfo{author}{Kim, J.}, \bibinfo{author}{Kuzyk, M.~C.},
  \bibinfo{author}{Han, K.}, \bibinfo{author}{Wang, H.} \&
  \bibinfo{author}{Bahl, G.}
\newblock \bibinfo{title}{Non-reciprocal brillouin scattering induced
  transparency}.
\newblock \emph{\bibinfo{journal}{{Nat. Phys.}}} \textbf{\bibinfo{volume}{11}},
  \bibinfo{pages}{275} (\bibinfo{year}{2015}).

\bibitem{Taravati_PRB_2017}
\bibinfo{author}{Taravati, S.}, \bibinfo{author}{Chamanara, N.} \&
  \bibinfo{author}{Caloz, C.}
\newblock \bibinfo{title}{Nonreciprocal electromagnetic scattering from a
  periodically space-time modulated slab and application to a quasisonic
  isolator}.
\newblock \emph{\bibinfo{journal}{{Phys. Rev. B}}}
  \textbf{\bibinfo{volume}{96}}, \bibinfo{pages}{165144}
  (\bibinfo{year}{2017}).

\bibitem{Taravati_PRB_SB_2017}
\bibinfo{author}{Taravati, S.}
\newblock \bibinfo{title}{Self-biased broadband magnet-free linear isolator
  based on one-way space-time coherency}.
\newblock \emph{\bibinfo{journal}{{Phys. Rev. B}}}
  \textbf{\bibinfo{volume}{96}}, \bibinfo{pages}{235150}
  (\bibinfo{year}{2017}).

\bibitem{Bahl_2018time}
\bibinfo{author}{Sohn, D.~B.}, \bibinfo{author}{Kim, S.} \&
  \bibinfo{author}{Bahl, G.}
\newblock \bibinfo{title}{Time-reversal symmetry breaking with acoustic pumping
  of nanophotonic circuits}.
\newblock \emph{\bibinfo{journal}{{Nat. Photonics}}}
  \textbf{\bibinfo{volume}{12}}, \bibinfo{pages}{91} (\bibinfo{year}{2018}).

\bibitem{correas2019plasmonic}
\bibinfo{author}{Correas-Serrano, D.}, \bibinfo{author}{Paul, N.} \&
  \bibinfo{author}{Gomez-Diaz, J.}
\newblock \bibinfo{title}{Plasmonic and photonic isolators based on the
  spatiotemporal modulation of graphene}.
\newblock In \emph{\bibinfo{booktitle}{Micro-and Nanotechnology Sensors,
  Systems, and Applications XI}}, vol. \bibinfo{volume}{10982},
  \bibinfo{pages}{109821I} (\bibinfo{organization}{International Society for
  Optics and Photonics}, \bibinfo{year}{2019}).

\bibitem{estep2014magnetic}
\bibinfo{author}{Estep, N.~A.}, \bibinfo{author}{Sounas, D.~L.},
  \bibinfo{author}{Soric, J.} \& \bibinfo{author}{Al{\`u}, A.}
\newblock \bibinfo{title}{Magnetic-free non-reciprocity and isolation based on
  parametrically modulated coupled-resonator loops}.
\newblock \emph{\bibinfo{journal}{Nat. Phys.}} \textbf{\bibinfo{volume}{10}},
  \bibinfo{pages}{923--927} (\bibinfo{year}{2014}).

\bibitem{reiskarimian2016magnetic}
\bibinfo{author}{Reiskarimian, N.} \& \bibinfo{author}{Krishnaswamy, H.}
\newblock \bibinfo{title}{Magnetic-free non-reciprocity based on staggered
  commutation}.
\newblock \emph{\bibinfo{journal}{Nat. Commun.}} \textbf{\bibinfo{volume}{7}}
  (\bibinfo{year}{2016}).

\bibitem{Fan_mats_2017}
\bibinfo{author}{Shi, Y.}, \bibinfo{author}{Han, S.} \& \bibinfo{author}{Fan,
  S.}
\newblock \bibinfo{title}{Optical circulation and isolation based on indirect
  photonic transitions of guided resonance modes}.
\newblock \emph{\bibinfo{journal}{{ACS Photonics}}}
  \textbf{\bibinfo{volume}{4}}, \bibinfo{pages}{1639--1645}
  (\bibinfo{year}{2017}).

\bibitem{Salary_2018}
\bibinfo{author}{Salary, M.~M.}, \bibinfo{author}{Jafar-Zanjani, S.} \&
  \bibinfo{author}{Mosallaei, H.}
\newblock \bibinfo{title}{Electrically tunable harmonics in time-modulated
  metasurfaces for wavefront engineering}.
\newblock \emph{\bibinfo{journal}{New J. Phys.}} \textbf{\bibinfo{volume}{20}},
  \bibinfo{pages}{123023} (\bibinfo{year}{2018}).

\bibitem{salary2019dynamically}
\bibinfo{author}{Salary, M.~M.}, \bibinfo{author}{Farazi, S.} \&
  \bibinfo{author}{Mosallaei, H.}
\newblock \bibinfo{title}{A dynamically modulated all-dielectric metasurface
  doublet for directional harmonic generation and manipulation in
  transmission}.
\newblock \emph{\bibinfo{journal}{{Adv. Opt. Mater.}}} \bibinfo{pages}{1900843}
  (\bibinfo{year}{2019}).

\bibitem{zang2019nonreciprocal_metas}
\bibinfo{author}{Zang, J.} \emph{et~al.}
\newblock \bibinfo{title}{Nonreciprocal wavefront engineering with
  time-modulated gradient metasurfaces}.
\newblock \emph{\bibinfo{journal}{{Phys. Rev. Appl.}}}
  \textbf{\bibinfo{volume}{11}}, \bibinfo{pages}{054054}
  (\bibinfo{year}{2019}).

\bibitem{Taravati_PRB_Mixer_2018}
\bibinfo{author}{Taravati, S.}
\newblock \bibinfo{title}{Aperiodic space-time modulation for pure frequency
  mixing}.
\newblock \emph{\bibinfo{journal}{{Phys. Rev. B}}}
  \textbf{\bibinfo{volume}{97}}, \bibinfo{pages}{115131}
  (\bibinfo{year}{2018}).

\bibitem{Taravati_LWA_2017}
\bibinfo{author}{Taravati, S.} \& \bibinfo{author}{Caloz, C.}
\newblock \bibinfo{title}{Mixer-duplexer-antenna leaky-wave system based on
  periodic space-time modulation}.
\newblock \emph{\bibinfo{journal}{{IEEE Trans. Antennas Propagat.}}}
  \textbf{\bibinfo{volume}{65}}, \bibinfo{pages}{442 -- 452}
  (\bibinfo{year}{2017}).

\bibitem{shanks1961new}
\bibinfo{author}{Shanks, H.}
\newblock \bibinfo{title}{A new technique for electronic scanning}.
\newblock \emph{\bibinfo{journal}{IEEE Trans. Antennas Propag.}}
  \textbf{\bibinfo{volume}{9}}, \bibinfo{pages}{162--166}
  (\bibinfo{year}{1961}).

\bibitem{Alu_PNAS_2016}
\bibinfo{author}{Hadad, Y.}, \bibinfo{author}{Soric, J.~C.} \&
  \bibinfo{author}{Al\`{u}, A.}
\newblock \bibinfo{title}{Breaking temporal symmetries for emission and
  absorption}.
\newblock \emph{\bibinfo{journal}{Proc. Natl. Acad. Sci.}}
  \textbf{\bibinfo{volume}{113}}, \bibinfo{pages}{3471--3475}
  (\bibinfo{year}{2016}).

\bibitem{ramaccia2018nonreciprocity}
\bibinfo{author}{Ramaccia, D.}, \bibinfo{author}{Sounas, D.~L.},
  \bibinfo{author}{Al{\`u}, A.}, \bibinfo{author}{Bilotti, F.} \&
  \bibinfo{author}{Toscano, A.}
\newblock \bibinfo{title}{Nonreciprocity in antenna radiation induced by
  space-time varying metamaterial cloaks}.
\newblock \emph{\bibinfo{journal}{{IEEE Antennas Wirel. Propagat. Lett.}}}
  \textbf{\bibinfo{volume}{17}}, \bibinfo{pages}{1968--1972}
  (\bibinfo{year}{2018}).

\bibitem{salary2019nonreciprocal}
\bibinfo{author}{Salary, M.~M.}, \bibinfo{author}{Jafar-Zanjani, S.} \&
  \bibinfo{author}{Mosallaei, H.}
\newblock \bibinfo{title}{Nonreciprocal optical links based on time-modulated
  nanoantenna arrays: Full-duplex communication}.
\newblock \emph{\bibinfo{journal}{{Phys. Rev. B}}}
  \textbf{\bibinfo{volume}{99}}, \bibinfo{pages}{045416}
  (\bibinfo{year}{2019}).

\bibitem{taravati2018space}
\bibinfo{author}{Taravati, S.} \& \bibinfo{author}{Kishk, A.~A.}
\newblock \bibinfo{title}{Space-time-varying surface-wave antenna}.
\newblock In \emph{\bibinfo{booktitle}{2018 18th International Symposium on
  Antenna Technology and Applied Electromagnetics (ANTEM)}}
  (\bibinfo{organization}{IEEE}, \bibinfo{year}{2018}).

\bibitem{taravati2019_mix_ant}
\bibinfo{author}{Taravati, S.} \& \bibinfo{author}{Eleftheriades, G.~V.}
\newblock \bibinfo{title}{Mixer-antenna medium}.
\newblock In \emph{\bibinfo{booktitle}{2019 13th International Congress on
  Artificial Materials for Novel Wave Phenomena (Metamaterials)}}
  (\bibinfo{organization}{IEEE}, \bibinfo{year}{2019}).

\bibitem{alvarez2019coupling}
\bibinfo{author}{Alvarez-Melcon, A.}, \bibinfo{author}{Wu, X.},
  \bibinfo{author}{Zang, J.}, \bibinfo{author}{Liu, X.} \&
  \bibinfo{author}{Gomez-Diaz, J.}
\newblock \bibinfo{title}{Coupling matrix representation of nonreciprocal
  filters based on time modulated resonators}.
\newblock \emph{\bibinfo{journal}{arXiv preprint arXiv:1905.08340}}
  (\bibinfo{year}{2019}).

\bibitem{wu2019isolating}
\bibinfo{author}{Wu, X.} \emph{et~al.}
\newblock \bibinfo{title}{Isolating bandpass filters using time-modulated
  resonators}.
\newblock \emph{\bibinfo{journal}{{IEEE Trans. Microw. Theory Techn.}}}
  \textbf{\bibinfo{volume}{67}}, \bibinfo{pages}{2331--2345}
  (\bibinfo{year}{2019}).

\bibitem{shlivinski2018beyond}
\bibinfo{author}{Shlivinski, A.} \& \bibinfo{author}{Hadad, Y.}
\newblock \bibinfo{title}{Beyond the bode-fano bound: Wideband impedance
  matching for short pulses using temporal switching of transmission-line
  parameters}.
\newblock \emph{\bibinfo{journal}{{Phys. Rev. Lett.}}}
  \textbf{\bibinfo{volume}{121}}, \bibinfo{pages}{204301}
  (\bibinfo{year}{2018}).

\bibitem{huidobro2019fresnel}
\bibinfo{author}{Huidobro, P.}, \bibinfo{author}{Galiffi, E.},
  \bibinfo{author}{Guenneau, S.}, \bibinfo{author}{Craster, R.} \&
  \bibinfo{author}{Pendry, J.}
\newblock \bibinfo{title}{Fresnel drag in space-time modulated metamaterials}.
\newblock \emph{\bibinfo{journal}{arXiv preprint arXiv:1908.05883}}
  (\bibinfo{year}{2019}).

\bibitem{zhang2018space}
\bibinfo{author}{Zhang, L.} \emph{et~al.}
\newblock \bibinfo{title}{Space-time-coding digital metasurfaces}.
\newblock \emph{\bibinfo{journal}{{Nat. Commun.}}}
  \textbf{\bibinfo{volume}{9}}, \bibinfo{pages}{4334} (\bibinfo{year}{2018}).

\bibitem{taravati_PRApp_2019}
\bibinfo{author}{Taravati, S.} \& \bibinfo{author}{Eleftheriades, G.~V.}
\newblock \bibinfo{title}{Generalized space-time periodic diffraction gratings:
  Theory and applications}.
\newblock \emph{\bibinfo{journal}{{Phys. Rev. Appl.}}}
  \textbf{\bibinfo{volume}{12}}, \bibinfo{pages}{024026}
  (\bibinfo{year}{2019}).

\end{thebibliography}

\end{document}